\documentclass[english,intlimits,longbibliography,nofootinbib,10pt]{revtex4-2}
\usepackage{lmodern}

\usepackage[T1]{fontenc}
\usepackage[latin9]{inputenc}
\usepackage{color}
\usepackage{babel}
\usepackage{amsmath}
\usepackage{amssymb}
\usepackage{graphicx}
\usepackage[unicode=true,pdfusetitle,
 bookmarks=true,bookmarksnumbered=false,bookmarksopen=false,
 breaklinks=false,pdfborder={0 0 0},pdfborderstyle={},backref=false,colorlinks=true]
 {hyperref}

\makeatletter
\usepackage{graphicx}
\usepackage{amsmath}
\hypersetup{colorlinks,allcolors=blue,breaklinks=true}

\makeatother

\begin{document}
\title{Extensive Long-Range Entanglement in a Nonequilibrium Steady State}
\author{Shachar Fraenkel}
\email{shacharf@mail.tau.ac.il}

\author{Moshe Goldstein}
\affiliation{Raymond and Beverly Sackler School of Physics and Astronomy, Tel Aviv
University, Tel Aviv 6997801, Israel}
\begin{abstract}
Entanglement measures constitute powerful tools in the quantitative
description of quantum many-body systems out of equilibrium. We study
entanglement in the current-carrying steady state of a paradigmatic
one-dimensional model of noninteracting fermions at zero temperature
in the presence of a scatterer. We show that disjoint intervals located
on opposite sides of the scatterer, and within similar distances from
it, maintain volume-law entanglement regardless of their separation,
as measured by their fermionic negativity and coherent information.
The mutual information of the intervals, which quantifies the total
correlations between them, follows a similar scaling. Interestingly,
this scaling entails in particular that if the position of one of
the intervals is kept fixed, then the correlation measures depend
non-monotonically on the distance between the intervals. By deriving
exact expressions for the extensive terms of these quantities, we
prove their simple functional dependence on the scattering probabilities,
and demonstrate that the strong long-range entanglement is generated
by the coherence between the transmitted and reflected parts of propagating
particles within the bias-voltage window. The generality and simplicity
of the model suggest that this behavior should characterize a large
class of nonequilibrium steady states.
\end{abstract}
\maketitle
\tableofcontents

\section{Introduction}

Within the broad field of quantum many-body physics, the study of
nonequilibrium phenomena is becoming increasingly intertwined with
the analysis of entanglement witnesses. In particular, the scaling
of various entanglement measures with the size of a subsystem quantitatively
captures canonical nonequilibrium behaviors, such as thermalization~\citep{doi:10.1080/00018732.2016.1198134,doi:10.1073/pnas.1703516114,Deutsch_2018}
or the violation thereof~\citep{doi:10.1146/annurev-conmatphys-031214-014726,RevModPhys.91.021001,Serbyn2021},
in closed systems subjected to an initial quench. In quench problems
of this type, transient effects of long-range entanglement are signatures
of integrability~\citep{Mesty_n_2017,10.21468/SciPostPhys.4.3.017,PhysRevB.100.115150,Alba_2019},
and the dynamics as well as the stationary values of the entanglement
entropy, negativity, and mutual information are used for the classification
of out-of-equilibrium models and their phases~\citep{Eisler_2012,PhysRevLett.111.127201,PhysRevLett.111.127205,PhysRevX.3.031015,PhysRevD.89.066015,PhysRevA.89.032321,Eisler_2014,Coser_2014,PhysRevB.92.075109,HOOGEVEEN201578,Gruber_2020,Paul2022Hidden}.

This success motivates the examination of entanglement properties
also in open systems, and specifically those of their steady states,
which may give rise to unique entanglement structures~\citep{PhysRevB.96.054302,PhysRevB.101.180301,PhysRevB.103.035108,PhysRevB.103.L020302,carollo2021emergent,10.21468/SciPostPhys.11.4.085,10.21468/SciPostPhys.12.1.011}.
Current-carrying states of inhomogeneous systems offer a promising
ground for such an analysis, as recent studies have revealed that
they can naturally sustain long-range quantum coherent correlations~\citep{PhysRevX.9.021007,PhysRevLett.123.080601,PhysRevX.13.011045}.
In this context, scaling laws of steady-state entanglement measures
were shown to be closely related to the localized-diffusive phase
transition of the open noninteracting Anderson model~\citep{PhysRevLett.123.110601}.
In this work we show that nonequilibrium conditions may lead to an
even more striking behavior of quantum information measures. This
is achieved through the study of an elementary model for an inhomogeneous
system in a current-carrying state, where the mechanism underlying
its unusual entanglement properties is exceptionally transparent.

Beyond the role of entanglement measures as fundamental quantities,
their estimation is inextricably linked to the construction of useful
tensor-network simulations of condensed matter systems~\citep{doi:10.1080/14789940801912366,SCHOLLWOCK201196}.
Strong (volume-law) entanglement, which is commonly found in nonequilibrium
quantum many-body states~\citep{PhysRevLett.111.127205,doi:10.1073/pnas.1703516114,PhysRevLett.124.137701},
impedes the utility of these simulation methods~\citep{PhysRevLett.100.030504}.
One possible key for their improvement is thus the uncovering of nontrivial
entanglement structures in certain classes of states, like the one
that is the subject of this work. Steady states that are predicted
to give rise to strong entanglement are also of potential interest
from a technological standpoint, as entanglement is an essential resource
for quantum information applications~\citep{PhysRevLett.70.1895,Bennett2000,RevModPhys.74.145}.

In this work, we examine the long-range entanglement induced by a
current-conserving scatterer in the voltage-biased steady state of
a 1D noninteracting fermion system. We treat this problem generally,
without imposing specific assumptions regarding the structure of the
scatterer other than it being smaller compared to all other length
scales. We study the correlations between two disjoint subsystems
located on opposite sides of the scatterer: $A_{{\scriptscriptstyle L}}$
on its left, and $A_{{\scriptscriptstyle R}}$ on its right. The total
amount of correlations is regularly quantified using the mutual information
(MI) between the two subsystems, 
\begin{equation}
{\cal I}_{A_{L}:A_{R}}={\cal S}_{A_{L}}+{\cal S}_{A_{R}}-{\cal S}_{A}.\label{eq:MI-definition}
\end{equation}
Here $A=A_{{\scriptscriptstyle L}}\cup A_{{\scriptscriptstyle R}}$,
and ${\cal S}_{X}=-{\rm Tr}\left[\rho_{{\scriptscriptstyle X}}\ln\rho_{{\scriptscriptstyle X}}\right]$
is the von Neumann entanglement entropy of a subsystem $X$~\citep{RevModPhys.81.865},
with $\rho_{{\scriptscriptstyle X}}$ being the reduced density matrix
of $X$.

Given that $A$ is in a mixed state, however, the MI has limitations
as a measure of entanglement, since it takes into account both classical
and quantum correlations~\citep{PhysRevA.72.032317}. Therefore,
we also address the fermionic negativity~\citep{PhysRevB.95.165101,PhysRevB.97.165123,PhysRevA.99.022310}
between $A_{{\scriptscriptstyle L}}$ and $A_{{\scriptscriptstyle R}}$,
an entanglement monotone defined as 
\begin{equation}
{\cal E}=\ln{\rm Tr}\sqrt{\left(\widetilde{\rho}_{{\scriptscriptstyle A}}\right)^{\dagger}\widetilde{\rho}_{{\scriptscriptstyle A}}},
\end{equation}
where $\widetilde{\rho}_{{\scriptscriptstyle A}}$ is obtained from
$\rho_{{\scriptscriptstyle A}}$ by applying a partial time-reversal
to either $A_{{\scriptscriptstyle L}}$ or $A_{{\scriptscriptstyle R}}$.
Interestingly, our analysis shows that the MI and negativity follow
a similar scaling, a scaling which to the best of our knowledge has
not been previously observed in a natural physical scenario.
\begin{figure*}
\begin{centering}
\includegraphics[viewport=25bp 35bp 1500bp 700bp,clip,width=1\textwidth]{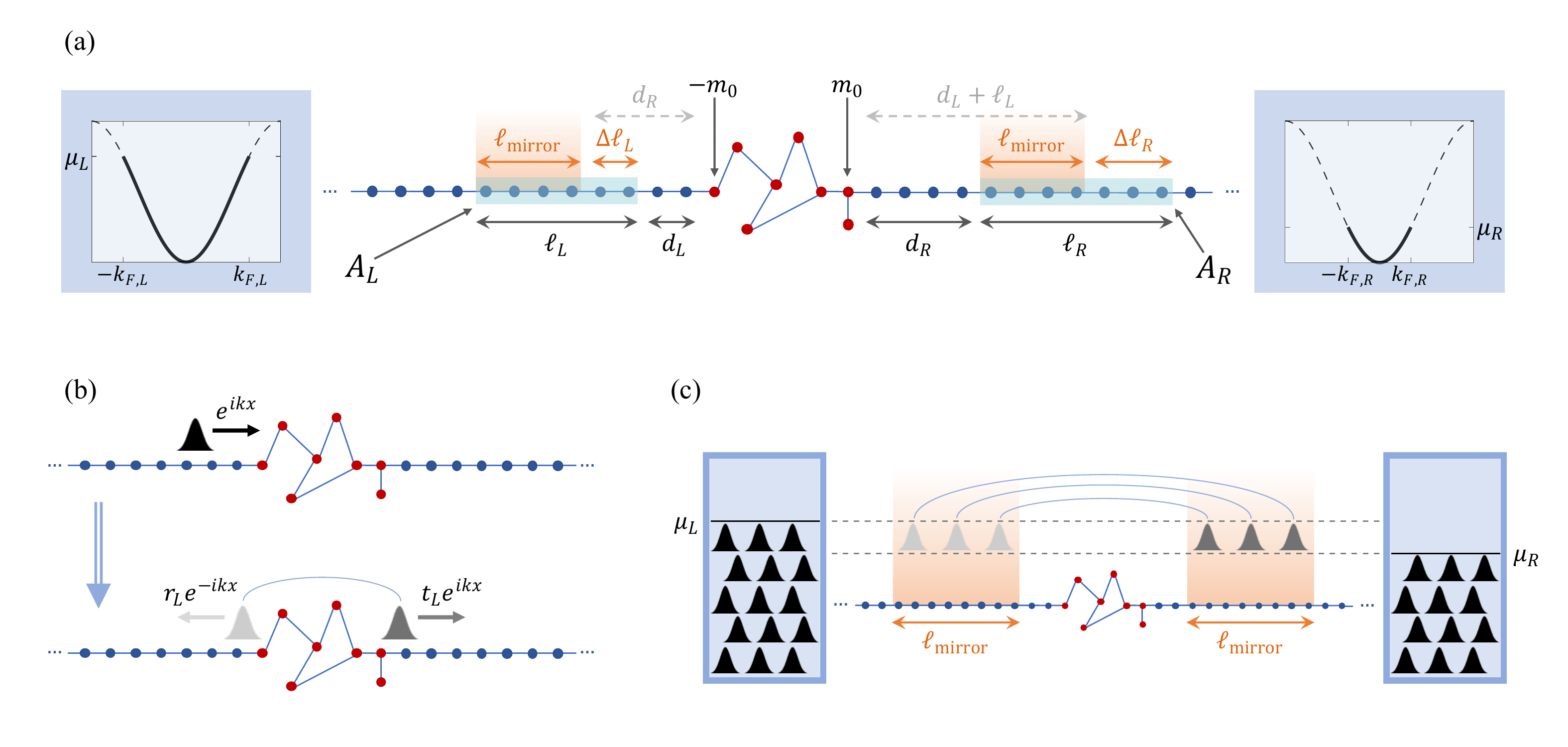}
\par\end{centering}
\caption{\label{fig:Model-sketch}(a) Schematic sketch of the model: Red circles
mark lattice sites in the scattering region, while sites outside this
region are marked in blue. Noninteracting reservoirs with different
chemical potentials are connected to the two ends of the chain. See
the text (Sec.~\ref{sec:Nonequilibrium-model}) for details regarding
the notations. Bottom panels: An intuitive picture for the origin
of the steady-state entanglement structure. (b) Any incoming wavepacket
(black) is split by the scattering region into a transmitted part
(dark gray) and a reflected part (light gray), with amplitudes determined
by the associated scattering matrix (see Eq.~(\ref{eq:Scattering_matrix})).
The transmitted and reflected parts are coherently correlated and
thus generate entanglement. (c) Split wavepackets with energies within
the voltage window strongly entangle regions that mirror each other
with respect to the position of the scattering region. Correlation
measures exhibit long-range volume-law scaling, since the number of
split wavepackets shared by these mirroring regions is proportional
to their length and independent of their spatial separation.}
\end{figure*}

As our main result, we find that both the MI and the negativity scale
linearly with $\ell_{{\rm mirror}}$, the number of sites in $A_{{\scriptscriptstyle L}}$
that, under reflection with respect to the position of the scatterer,
overlap with sites in $A_{{\scriptscriptstyle R}}$ (see Fig.~\ref{fig:Model-sketch}(a)
for an illustration). Remarkably, this steady-state extensive entanglement
is long-ranged, as the observed volume-law scaling does not decay
with the (similar) distance of the mirroring sites from the scatterer.
Moreover, the entanglement depends non-monotonically on the distance
of either $A_{{\scriptscriptstyle L}}$ or $A_{{\scriptscriptstyle R}}$
from the scatterer. We analytically derive exact formulas for the
asymptotic scaling of the MI and the negativity (Eqs.~(\ref{eq:MI_volume_law})--(\ref{eq:Negativity_volume_law})).
Additionally, we demonstrate that the coherent information (CI)~\citep{PhysRevA.54.2629,Horodecki2005},
\begin{equation}
I\!\left(A_{{\scriptscriptstyle L}}\rangle A_{{\scriptscriptstyle R}}\right)={\cal S}_{A_{R}}-{\cal S}_{A},\label{eq:CI-definition}
\end{equation}
is not only positive (which is impossible classically) when $\ell_{{\rm mirror}}$
is large enough, but also grows with $\ell_{{\rm mirror}}$ according
to a volume law (Eq.~(\ref{eq:CI_volume_law}))\footnote{The definition of the CI is evidently not symmetric with respect to
the two subsystems $A_{{\scriptscriptstyle L}}$ and $A_{{\scriptscriptstyle R}}$.
Our choice to examine $I\!\left(A_{{\scriptscriptstyle L}}\rangle A_{{\scriptscriptstyle R}}\right)$
rather than $I\!\left(A_{{\scriptscriptstyle R}}\rangle A_{{\scriptscriptstyle L}}\right)$
is arbitrary, and we maintain this choice throughout the text for
convenience.}. The CI is a lower bound to the squashed entanglement~\citep{doi:10.1063/1.1643788,Carlen2012},
another rigorous entanglement measure with axiomatically desirable
properties~\citep{RevModPhys.81.865}, which therefore obeys an extensive
scaling as well in regimes where $I\!\left(A_{{\scriptscriptstyle L}}\rangle A_{{\scriptscriptstyle R}}\right)>0$.

A simple intuitive explanation for these results is provided by considering
that the scatterer splits each incoming single-particle wavepacket
into two coherently correlated counter-propagating parts (Fig.~\ref{fig:Model-sketch}(b)).
Each such split wavepacket with energy within the voltage window generates
entanglement, since detecting the particle in one subsystem prohibits
its presence in the other. As the number of such wavepackets is proportional
to $\ell_{{\rm mirror}}$ and independent of the distance between
the subsystems (Fig.~\ref{fig:Model-sketch}(c)), the correlation
measures exhibit a similar behavior.

The paper is organized as follows. In Sec.~\ref{sec:Nonequilibrium-model}
we introduce the model for the system and its nonequilibrium steady
state that are the subject of this work. In Sec.~\ref{sec:Asymptotics-of-correlation-measures}
we report our analytical results for correlation measures in the steady
state. We point out the salient features of these results, and support
them through comparisons to numerical results (computed for a specific
choice of the scatterer). Sec.~\ref{sec:Analytical-method} outlines
the derivation of the analytical results, and is limited to the conceptually
crucial steps in the derivation, while the technical aspects of the
process are mostly discussed in the appendices. In Sec.~\ref{sec:Discussion-and-outlook}
we conclude and mention potential future directions arising from this
work.

Additionally, the paper includes four technical appendices. In Appendix
\ref{sec:Two-point-correlations} we derive the two-point correlation
function, which served as the basic ingredient in all of our calculations.
Appendix \ref{sec:Calculation-of-Renyi-entropies} presents the technical
details of our computation method for subsystem entropies, from which
(as explained in Sec.~\ref{sec:Analytical-method}) the asymptotics
of the MI and CI can be immediately derived. Appendix \ref{sec:Calculation-of-negativity}
summarizes the derivation of the formula for the fermionic negativity,
which is based on the same method. Finally, in Appendix \ref{sec:Additional-numerical-tests}
we complement the numerical results included in Sec.~\ref{sec:Asymptotics-of-correlation-measures}
with additional numerical tests corroborating our analysis.

\section{Nonequilibrium model\label{sec:Nonequilibrium-model}}

We consider a 1D lattice, occupied by noninteracting fermions and
connected at its ends to two reservoirs with different chemical potentials,
$\mu_{{\scriptscriptstyle L}}\neq\mu_{{\scriptscriptstyle R}}$, at
zero temperature. The lattice is assumed to be of infinite length,
and it is modeled as a tight-binding chain that is homogeneous everywhere,
except for a small region at the center of the chain, which we dub
\emph{the scattering region}. The Hamiltonian is thus of the form
\begin{align}
{\cal H} & =-\eta\!\sum_{m=m_{0}}^{\infty}\!\!\left[c_{m}^{\dagger}c_{m+1}+c_{-m}^{\dagger}c_{-m-1}+{\rm h.c.}\right]+{\cal H}_{{\rm scat}}.\label{eq:Model-Hamiltonian}
\end{align}
Here $c_{m}$ is a fermionic annihilation operator for the $m$th
lattice site, $\eta>0$ is a hopping amplitude, $m=\pm m_{0}$ designate
the boundaries of the scattering region, and ${\cal H}_{{\rm scat}}$
pertains only to sites with $\left|m\right|\le m_{0}$ and breaks
the homogeneity, e.g., through modified hopping terms, on-site energies,
or side-attached sites.

The scattering region can be associated with a $2\times2$ unitary
scattering matrix~\citep{merzbacher1998quantum}, defined for any
lattice momentum $0<k<\pi$:
\begin{equation}
S\!\left(k\right)=\left(\begin{array}{cc}
r_{{\scriptscriptstyle L}}\!\left(k\right) & t_{{\scriptscriptstyle R}}\!\left(k\right)\\
t_{{\scriptscriptstyle L}}\!\left(k\right) & r_{{\scriptscriptstyle R}}\!\left(k\right)
\end{array}\right).\label{eq:Scattering_matrix}
\end{equation}
The diagonal (off-diagonal) entries of this matrix stand for reflection
(transmission) amplitudes; the left (right) column contains the scattering
amplitudes for a particle originating in the left (right) reservoir
with momentum $k>0$ ($-k<0$). The squared moduli of the entries
correspond to the transmission and reflection probabilities, respectively
${\cal T}\!\left(\left|k\right|\right)$ and ${\cal R}\!\left(\left|k\right|\right)=1-{\cal T}\!\left(\left|k\right|\right)$,
for a particle originating in either reservoir with momentum $k$.
These scattering probabilities are the sole property of the scatterer
on which our analytical results depend.

The single-particle eigenbasis of the Hamiltonian is comprised of
extended scattering states with energies $\varepsilon=-2\eta\cos k$,
and of bound states localized near the scattering region~\citep{merzbacher1998quantum,doi:10.1063/1.525968};
we ignore the latter in our analysis, as they contribute negligibly
to correlations between sites that are distant from the scatterer.
The current-carrying many-body steady state is pure, with single-particle
scattering states originating in the left (right) reservoir occupied
up to a Fermi momentum $k_{{\scriptscriptstyle F,L}}>0$ ($-k_{{\scriptscriptstyle F,R}}<0$),
as shown schematically in Fig.~\ref{fig:Model-sketch}(a). The Fermi
momenta are related to the chemical potentials through $\mu_{i}=-2\eta\cos k_{{\scriptscriptstyle F,i}}$
($i=L,R$).

Correlation and entanglement measures are calculated with respect
to two subsystems $A_{{\scriptscriptstyle L}}$ and $A_{{\scriptscriptstyle R}}$,
each comprised of contiguous sites, with lengths $\ell_{i}$ and distances
$d_{i}\ge0$ ($i=L,R$) from the scattering region (all of which are
assumed to be much larger than the size of the scattering region,
$2m_{0}+1$): $A_{{\scriptscriptstyle L}}$ contains the sites $m$
such that $-d_{{\scriptscriptstyle L}}-\ell_{{\scriptscriptstyle L}}\le m+m_{0}\le-d_{{\scriptscriptstyle L}}-1$,
while $A_{{\scriptscriptstyle R}}$ contains the sites $m$ such that
$d_{{\scriptscriptstyle R}}+1\le m-m_{0}\le d_{{\scriptscriptstyle R}}+\ell_{{\scriptscriptstyle R}}$
(see Fig.~\ref{fig:Model-sketch}(a)). $\ell_{{\rm mirror}}=\max\left\{ \min\left\{ d_{{\scriptscriptstyle L}}+\ell_{{\scriptscriptstyle L}},d_{{\scriptscriptstyle R}}+\ell_{{\scriptscriptstyle R}}\right\} -\max\left\{ d_{{\scriptscriptstyle L}},d_{{\scriptscriptstyle R}}\right\} ,0\right\} $
denotes the number of mirroring pairs $\left(-m,m\right)\in A_{{\scriptscriptstyle L}}\times A_{{\scriptscriptstyle R}}$,
and we also define $\Delta\ell_{i}=\ell_{i}-\ell_{{\rm mirror}}$.

\section{Asymptotics of correlation measures\label{sec:Asymptotics-of-correlation-measures}}

The leading behaviors of the MI and the negativity can be encapsulated
by that of the R\'enyi MI, defined as 
\begin{equation}
{\cal I}_{A_{L}:A_{R}}^{\left(n\right)}=S_{A_{L}}^{\left(n\right)}+S_{A_{R}}^{\left(n\right)}-S_{A}^{\left(n\right)},
\end{equation}
where $S_{X}^{\left(n\right)}=\frac{1}{1-n}\ln{\rm Tr}\left[\left(\rho_{{\scriptscriptstyle X}}\right)^{n}\right]$
are R\'enyi entropies (which converge to ${\cal S}_{X}$ as $n\to1$).
We report that, for the nonequilibrium steady state described above,
the R\'enyi MI follows a volume-law scaling with $\ell_{{\rm mirror}}$,
\begin{equation}
{\cal I}_{A_{L}:A_{R}}^{\left(n\right)}\sim\frac{\ell_{{\rm mirror}}}{1-n}\!\int_{k_{-}}^{k_{+}}\frac{dk}{\pi}\ln\!\left[\left({\cal T}\!\left(k\right)\right)^{n}+\left({\cal R}\!\left(k\right)\right)^{n}\right]+\ldots,\label{eq:Renyi_MI_volume_law}
\end{equation}
where $k_{-}=\min\!\left\{ k_{{\scriptscriptstyle F,L}},k_{{\scriptscriptstyle F,R}}\right\} $
and $k_{+}=\max\!\left\{ k_{{\scriptscriptstyle F,L}},k_{{\scriptscriptstyle F,R}}\right\} $
are the two Fermi momenta that bound the voltage window. The ellipsis
(which will be henceforth omitted) represents subleading terms, the
dominant of which are logarithmic in the different length scales ($\ell_{i}$,
$d_{i}$, and combinations thereof). Full exact expressions for these
logarithmic terms can be obtained in the long-range limit $d_{i}/\ell_{i}\to\infty$
(with $d_{{\scriptscriptstyle L}}-d_{{\scriptscriptstyle R}}$ kept
fixed) using methods related to the asymptotic calculation of Toeplitz
determinants~\citep{Jin2004,10.2307/23030524,PhysRevA.92.042334,PhysRevA.97.062301};
the results for these subleading corrections will be discussed in
a separate publication~\citep{FuturePublication}.

The MI is related to the R\'enyi MI simply by its definition, through
the equality ${\cal I}_{A_{L}:A_{R}}=\lim_{n\to1}{\cal I}_{A_{L}:A_{R}}^{\left(n\right)}$,
yielding the following asymptotics:
\begin{equation}
{\cal I}_{A_{L}:A_{R}}\sim\ell_{{\rm mirror}}\!\int_{k_{-}}^{k_{+}}\frac{dk}{\pi}\left[-{\cal T}\ln\!{\cal T}\!-\!{\cal R}\ln\!{\cal R}\right].\label{eq:MI_volume_law}
\end{equation}
The negativity, on the other hand, is not \emph{a priori} directly
related to the R\'enyi MI, yet our analysis shows that, at the leading
(linear) order,
\begin{equation}
{\cal E}\sim\ell_{{\rm mirror}}\!\int_{k_{-}}^{k_{+}}\frac{dk}{\pi}\ln\!\left[{\cal T}^{1/2}+{\cal R}^{1/2}\right]\sim\frac{1}{2}{\cal I}_{A_{L}:A_{R}}^{\left(1/2\right)}.\label{eq:Negativity_volume_law}
\end{equation}
 The equality ${\cal I}_{A_{L}:A_{R}}^{\left(1/2\right)}=2{\cal E}$
is known to arise in quenches of integrable systems~\citep{Alba_2019}.
Eqs.~(\ref{eq:MI_volume_law}) and (\ref{eq:Negativity_volume_law})
state that, for a generic non-trivial scatterer (i.e., unless ${\cal T}\!\left(k\right)\in\left\{ 0,1\right\} $
for all $k\in\left[k_{-},k_{+}\right]$), the MI and negativity both
exhibit extensive scaling with $\ell_{{\rm mirror}}$. Additionally,
we have found that the CI scales at the leading order as
\begin{equation}
I\!\left(A_{{\scriptscriptstyle L}}\rangle A_{{\scriptscriptstyle R}}\right)\sim\left(\ell_{{\rm mirror}}\!-\!\Delta\ell_{{\scriptscriptstyle L}}\right)\!\int_{k_{-}}^{k_{+}}\frac{dk}{2\pi}\left[-{\cal T}\ln\!{\cal T}\!-\!{\cal R}\ln\!{\cal R}\right],\label{eq:CI_volume_law}
\end{equation}
and so it grows linearly with $\ell_{{\rm mirror}}$ if $\Delta\ell_{{\scriptscriptstyle L}}$
is fixed. Crucially, the asymptotics in Eqs.~(\ref{eq:Renyi_MI_volume_law})--(\ref{eq:CI_volume_law})
do not depend on the magnitudes of $d_{i}$, and they hold even when
$d_{i}\gg\ell_{i}$. That is, the extensive entanglement is long-ranged,
and it holds even for subsystems that are very distant relative to
their lengths, but that still share mirroring sites. Eqs.~(\ref{eq:MI_volume_law})--(\ref{eq:CI_volume_law})
are the central results of this work.

\begin{figure}
\begin{centering}
\includegraphics[viewport=0bp 40bp 461bp 688bp,clip,width=0.5\columnwidth]{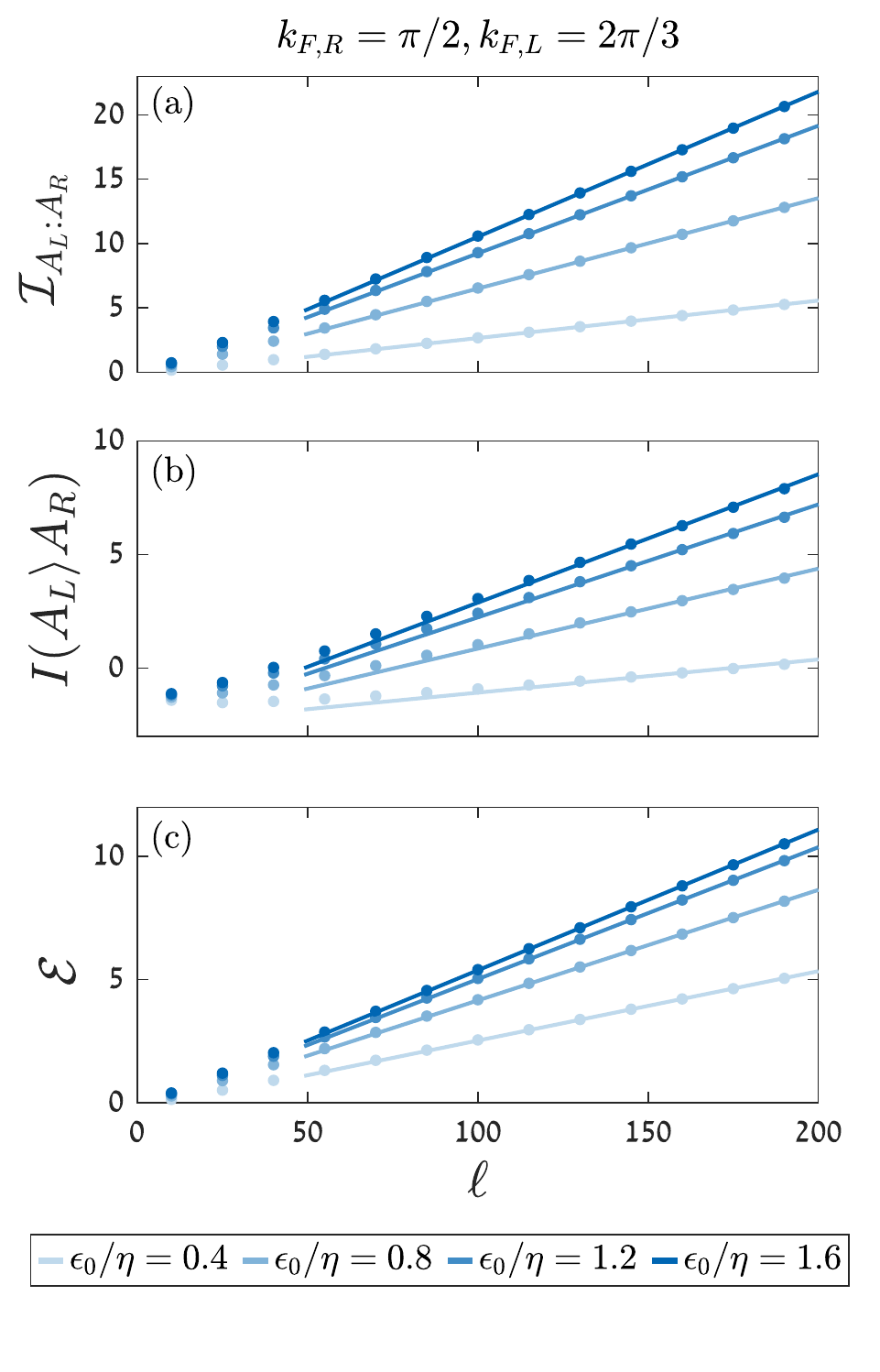}
\par\end{centering}
\caption{\label{fig:Num-Comparison-Symmetric}The single impurity model: Scaling
of (a) the mutual information, (b) the coherent information, and (c)
the fermionic negativity between subsystems $A_{{\scriptscriptstyle L}}$
and $A_{{\scriptscriptstyle R}}$ for the symmetric case $\ell_{{\scriptscriptstyle L}}=\ell_{{\scriptscriptstyle R}}=\ell$
and $d_{{\scriptscriptstyle L}}=d_{{\scriptscriptstyle R}}$, in the
limit $d_{i}\gg\ell_{i}$. The analytical results of Eqs.~(\ref{eq:MI_volume_law})--(\ref{eq:CI_volume_law})
for $\ell\ge50$ (lines) are compared to numerical results (dots)
for different values of the impurity energy $\epsilon_{0}$, with
the Fermi momenta fixed at $k_{{\scriptscriptstyle F,R}}=\pi/2$ and
$k_{{\scriptscriptstyle F,L}}=2\pi/3$.}
\end{figure}

The special symmetric case where $\ell_{{\scriptscriptstyle L}}=\ell_{{\scriptscriptstyle R}}=\ell$
and the subsystems are positioned symmetrically relative to the scatterer
($d_{{\scriptscriptstyle L}}=d_{{\scriptscriptstyle R}}$) is particularly
illuminating with regard to the nature of the steady-state entanglement.
In this case we have found that ${\cal S}_{A}$ scales sublinearly
with $\ell$, i.e.~$\lim_{\ell\to\infty}{\cal S}_{A}/\ell=0$ (see
Eq.~(\ref{eq:Renyi-entropies-asymptotics})). The combined subsystem
$A$ is therefore weakly entangled to the rest of the system, while
its two components -- one being the mirror image of the other --
maintain strong entanglement between them.

The volume-law terms in Eqs.~(\ref{eq:Renyi_MI_volume_law})--(\ref{eq:CI_volume_law})
are evidently generated by the scattering states within the voltage
window, with the contribution of each state in Eq.~(\ref{eq:Renyi_MI_volume_law})
being the equivalent of the statistical moment of its corresponding
transmission probability. This simple form allows to deduce that the
source of the long-range entanglement is the coherence between the
reflected part and the transmitted part of each scattered particle,
which arrive simultaneously at mirroring sites. In the steady state,
the constant particle current renders this strong entanglement a stationary
property, and the length scale $\ell_{{\rm mirror}}$ determines the
amount of entanglement as it is proportional to the number of scattered
particles shared by the two subsystems. The voltage bias and the non-trivial
scattering constitute necessary and generically-sufficient conditions
for the extensive terms in Eqs.~(\ref{eq:Renyi_MI_volume_law})--(\ref{eq:CI_volume_law})
to not vanish.

\begin{figure}
\begin{centering}
\includegraphics[viewport=0bp 10bp 460bp 620bp,clip,width=0.5\columnwidth]{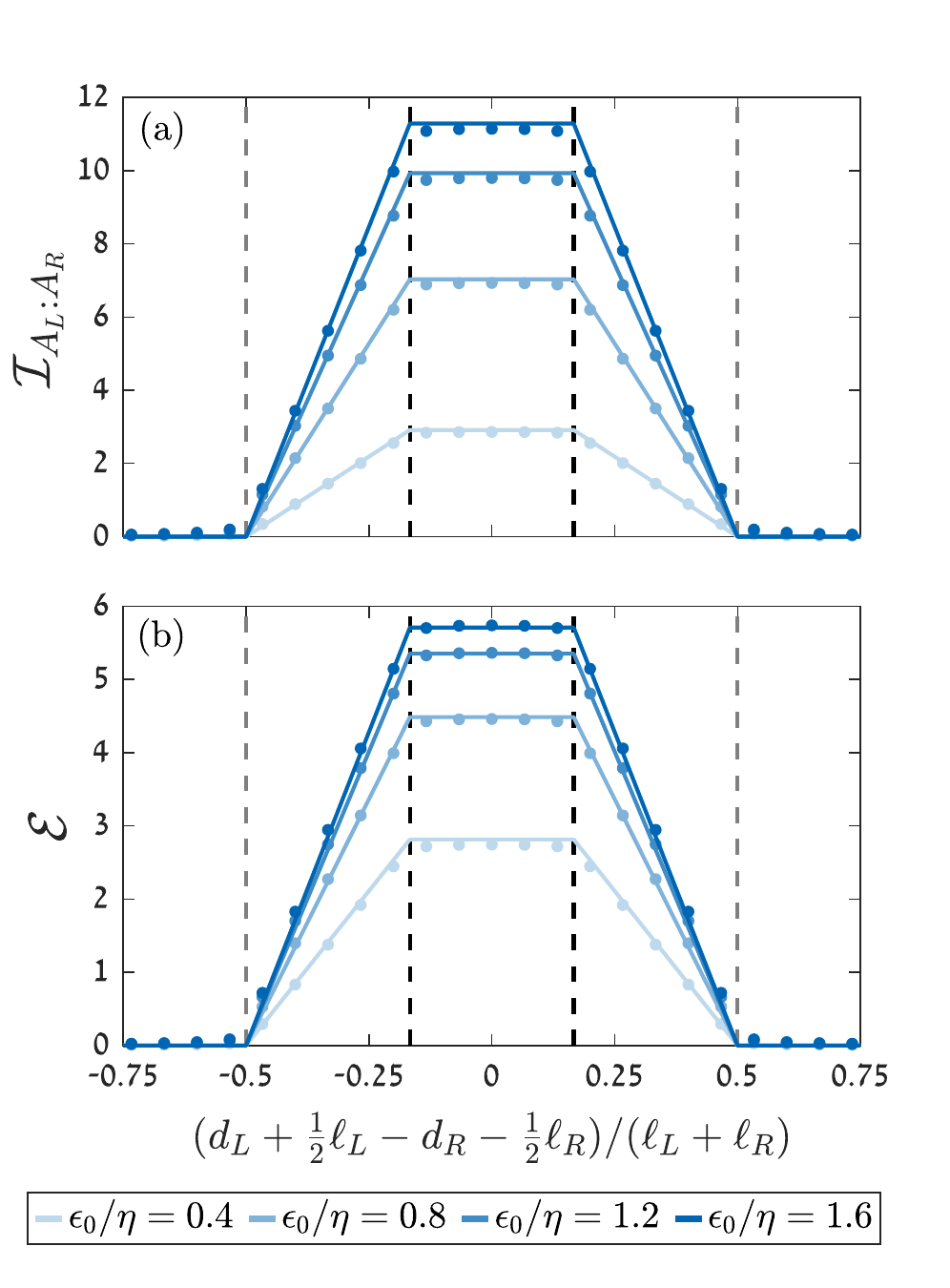}
\par\end{centering}
\caption{\label{fig:MI-Varying-Distance}The single impurity model: (a) The
mutual information and (b) the fermionic negativity between subsystems
$A_{{\scriptscriptstyle L}}$ and $A_{{\scriptscriptstyle R}}$ as
a function of their positions relative to the impurity. We fix $\ell_{{\scriptscriptstyle L}}=100$
and $\ell_{{\scriptscriptstyle R}}=200$, and observe the dependence
on $d_{{\scriptscriptstyle L}}-d_{{\scriptscriptstyle R}}$ in the
limit $d_{i}\gg\ell_{i}$. Analytical results (lines) are compared
to numerical results (dots) for different values of the impurity energy
$\epsilon_{0}$, with the Fermi momenta fixed at $k_{{\scriptscriptstyle F,R}}=\pi/2$
and $k_{{\scriptscriptstyle F,L}}=2\pi/3$. Letting $\bar{A}_{{\scriptscriptstyle L}}=\left\{ m|-\!m\in A_{{\scriptscriptstyle L}}\right\} $
denote the mirror image of $A_{{\scriptscriptstyle L}}$, black dashed
vertical lines mark the boundaries of the domain where $\bar{A}_{{\scriptscriptstyle L}}\subset A_{{\scriptscriptstyle R}}$,
while gray dashed vertical lines mark the boundaries of the domain
where $\bar{A}_{{\scriptscriptstyle L}}\cap A_{{\scriptscriptstyle R}}\protect\neq\phi$.}
\end{figure}

To support our general analytical results, we compared them to numerics
for a specific model where the scattering is a result of a single
impurity at the site $m=0$\footnote{The numerical results presented in Figs.~\ref{fig:Num-Comparison-Symmetric}
and \ref{fig:MI-Varying-Distance} were calculated in the limit $d_{i}/\ell_{i}\to\infty$,
which allows to simplify the expressions for the elements of two-point
correlation matrices, as explained in Appendix~\ref{sec:Two-point-correlations}.
In Appendix~\ref{sec:Additional-numerical-tests} we also compare
these numerical results to those computed for finite $d_{i}/\ell_{i}$,
demonstrating convergence for $d_{i}/\ell_{i}\gg1$.}. For this model, $m_{0}=0$ and ${\cal H}_{{\rm scat}}=\epsilon_{0}c_{0}^{\dagger}c_{0}$
in Eq.~(\ref{eq:Model-Hamiltonian}), $\epsilon_{0}$ being the impurity
energy. The scattering matrix for this model yields the transmission
probability
\begin{equation}
{\cal T}\!\left(k\right)=\frac{\sin^{2}k}{\sin^{2}k+\left(\epsilon_{0}/2\eta\right)^{2}}.
\end{equation}
Good agreement with numerics is manifest in Fig.~\ref{fig:Num-Comparison-Symmetric},
where, focusing on the aforementioned symmetric case with two intervals
of length $\ell$, we plotted the scaling with $\ell$ of all three
correlation measures for different ratios of $\epsilon_{0}/\eta$.
The analytical results of Eqs.~(\ref{eq:MI_volume_law})--(\ref{eq:CI_volume_law})
are plotted for $\ell\ge50$ (with a constant-in-$\ell$ additive
correction term as the only fitting parameter), as for small values
of $\ell$ there is a considerable contribution from subleading terms
beyond the leading volume-law term (an exact analytical result for
the most dominant subleading term, which is logarithmic in $\ell$,
is derived in Ref\@.~\citep{FuturePublication}).

In Fig.~\ref{fig:MI-Varying-Distance} we illustrate a rather counter-intuitive
attribute of our results, using the example of the single impurity
model. For fixed values of $\ell_{{\scriptscriptstyle L}}$ and $\ell_{{\scriptscriptstyle R}}$,
we plot the dependence of the MI and the negativity on the positions
of the subsystems, and observe that this dependence is non-monotonic.
Indeed, Eqs.~(\ref{eq:MI_volume_law})--(\ref{eq:CI_volume_law})
state that the long-range correlations are the strongest when the
overlap between one subsystem and the mirror image of the other is
maximal; if one subsystem is then brought closer to the other, this
overlap is reduced and so are the correlations. Fig.~\ref{fig:MI-Varying-Distance}
again showcases the good agreement of our analytical results with
numerics; the apparent slight deviations can be resolved once logarithmic
corrections are accounted for~\citep{FuturePublication}.

\section{Analytical method\label{sec:Analytical-method}}

This section delineates the main steps in the derivation of Eqs.~(\ref{eq:MI_volume_law})--(\ref{eq:CI_volume_law}),
while the discussion of the various technical steps is deferred to
the appendices. Subsec.~\ref{subsec:Derivation_of_MI_CI} focuses
on the derivation of the formulae for the MI and CI (both of which
are combinations of subsystem entropies), while Subsec.~\ref{subsec:Derivation_of_negativity}
deals with the derivation of the negativity asymptotics.

The joint starting point of these computations is the two-point correlation
function $\left\langle c_{j}^{\dagger}c_{m}\right\rangle $ for $j,m\in A$.
The absence of interactions entails that the states of the total system
and its subsystems are Gaussian, and thus entanglement is fully encoded
in two-point correlations~\citep{Peschel_2003,PhysRevB.95.165101,Shapourian_2019,PhysRevB.97.165123}.
The correlation function is given explicitly by
\begin{equation}
\left\langle c_{j}^{\dagger}c_{m}\right\rangle =\int_{-k_{F,R}}^{k_{F,L}}\frac{dk}{2\pi}\,u_{j}^{*}\!\left(k\right)u_{m}\!\left(k\right),\label{eq:Correlation_function_integral}
\end{equation}
where $u_{m}\!\left(k\right)$ is the (unnormalized) single-particle
wavefunction amplitude at site $m$ of the scattering state associated
with momentum $k$; namely, 
\begin{align}
u_{m>m_{0}}\!\left(k\right) & =\begin{cases}
e^{ikm}+r_{{\scriptscriptstyle R}}\!\left(\left|k\right|\right)e^{-ikm} & k<0,\\
t_{{\scriptscriptstyle L}}\!\left(\left|k\right|\right)e^{ikm} & k>0,
\end{cases}\nonumber \\
u_{m<-m_{0}}\!\left(k\right) & =\begin{cases}
t_{{\scriptscriptstyle R}}\!\left(\left|k\right|\right)e^{ikm} & k<0,\\
e^{ikm}+r_{{\scriptscriptstyle L}}\!\left(\left|k\right|\right)e^{-ikm} & k>0.
\end{cases}\label{eq:Unnormalized-wavefunctions}
\end{align}
Eq.~(\ref{eq:Correlation_function_integral}) is derived in Appendix~\ref{sec:Two-point-correlations},
where we also discuss how this expression may be simplified when $d_{i}\gg\ell_{i}$.

\subsection{Mutual information and coherent information\label{subsec:Derivation_of_MI_CI}}

The exact asymptotics of the MI and of the CI in Eqs.~(\ref{eq:MI_volume_law})
and (\ref{eq:CI_volume_law}) were obtained through the computation
of the R\'enyi entropies of $A_{{\scriptscriptstyle L}}$, $A_{{\scriptscriptstyle R}}$
and $A$. We describe here the main components of this computation,
referring the interested reader to Appendix~\ref{sec:Calculation-of-Renyi-entropies}
for full details.

Within the Gaussian steady state, the R\'enyi entropies of a subsystem
$X$ are reduced to functions of $C_{X}$, the correlation matrix
restricted to $X$ ($\left(C_{X}\right)_{jm}=\left\langle c_{j}^{\dagger}c_{m}\right\rangle $
where $j,m\in X$)~\citep{Peschel_2003}. Furthermore, these functions
admit a simple series expansion, on which our derivation relied. Namely,
the R\'enyi entropies are given by
\begin{align}
S_{X}^{\left(n\right)} & =\frac{1}{1-n}{\rm Tr}\ln\!\left[\left(C_{X}\right)^{n}+\left(\mathbb{I}-C_{X}\right)^{n}\right]\nonumber \\
 & =\frac{1}{1-n}\sum_{s=1}^{\infty}\frac{\left(-1\right)^{s+1}}{s}{\rm Tr}\!\left[\left\{ \left(C_{X}\right)^{n}+\left(\mathbb{I}-C_{X}\right)^{n}-\mathbb{I}\right\} ^{s}\right]\!.\label{eq:Renyi-from-correlations}
\end{align}
To obtain an analytical expression for the R\'enyi entropies, it is
therefore sufficient to calculate a general expression for moments
${\rm Tr}\!\left[\left(C_{X}\right)^{p}\right]$, with $p$ being
positive integers. Each such moment can be expressed in the form of
a $p$-dimensional integral,
\begin{equation}
{\rm Tr}\left[\left(C_{X}\right)^{p}\right]=\!\int_{\left[-k_{{\scriptscriptstyle F,R}},k_{{\scriptscriptstyle F,L}}\right]^{p}}\!\!\frac{d^{p}k}{\left(2\pi\right)^{p}}\prod_{j=1}^{p}\left[\sum_{m\in X}\!u_{m}\!\left(k_{j-1}\right)u_{m}^{*}\!\left(k_{j}\right)\right],\label{eq:Correlation-Moments}
\end{equation}
where we defined $k_{0}=k_{p}$. Each sum over $m\in X$ in Eq.~(\ref{eq:Correlation-Moments})
can be rewritten as an integral over a fictitious variable $\xi_{j}\in\left[-1,1\right]$,
such that ${\rm Tr}\left[\left(C_{X}\right)^{p}\right]$ is then expressed
as a $2p$-dimensional integral.

The specific form of this integral depends on the choice of $X$,
and is relatively involved (see Appendix~\ref{sec:Calculation-of-Renyi-entropies}).
We illustrate schematically the way forward by considering the case
of the connected subsystems $X=A_{i}$. For each of these subsystems,
Eq.~(\ref{eq:Correlation-Moments}) can be cast in the general form
\begin{equation}
{\rm Tr}\!\left[\left(C_{A_{i}}\right)^{p}\right]=\left(\frac{\ell_{i}}{4\pi}\right)^{p}\negthinspace\!\!\sum_{\overrightarrow{\tau}\!,\!\overrightarrow{\sigma}\in\left\{ 0,1\right\} ^{\otimes p}}\,\,\int_{\left[-k_{{\scriptscriptstyle F,R}},k_{{\scriptscriptstyle F,L}}\right]^{p}}\!\!\!\!\!\!d^{p}k\int_{\left[-1,1\right]^{p}}\!\!\!\!\!\!d^{p}\xi\,f_{\overrightarrow{\tau}\!,\!\overrightarrow{\sigma}}\!\left(\overrightarrow{k}\right)\exp\!\!\left[i\frac{\ell_{i}}{2}\sum_{j=1}^{p}\left(k_{\tau_{j-1}}-k_{\sigma_{j}}\right)\left(\xi_{j}+1\right)\right],\label{eq:Correlation-Moments-Fully-Integral}
\end{equation}
where $k_{\sigma_{j}}=\left(-1\right)^{\sigma_{j}}k_{j}$, and where
the functions $f_{\overrightarrow{\tau}\!,\!\overrightarrow{\sigma}}$
vanish for $\overrightarrow{k}\notin\left[-k_{{\scriptscriptstyle F,R}},k_{{\scriptscriptstyle F,L}}\right]^{p}$
and are independent of $\ell_{i}$. The origin of the exponential
term in Eq.~(\ref{eq:Correlation-Moments-Fully-Integral}) can be
traced back to the explicit forms of the wavefunctions $u_{m}\!\left(k\right)$
in Eq.~(\ref{eq:Unnormalized-wavefunctions}), which are superpositions
of $e^{\pm ikm}$.

An integral of the form of Eq.~(\ref{eq:Correlation-Moments-Fully-Integral})
admits a stationary phase approximation in the large-$\ell_{i}$ limit~\citep{doi:10.1137/1.9780898719260,PhysRevA.78.010306,10.21468/SciPostPhys.12.1.011};
leading-order contributions come only from summands with $\overrightarrow{\tau}=\overrightarrow{\sigma}$,
with the asymptotics of Eq.~(\ref{eq:Correlation-Moments-Fully-Integral})
given by
\begin{equation}
{\rm Tr}\!\left[\left(C_{A_{i}}\right)^{p}\right]\sim\ell_{i}\!\!\int_{-k_{F,R}}^{k_{F,L}}\!\!\frac{dk_{p}}{2\pi}\sum_{\overrightarrow{\sigma}\in\left\{ 0,1\right\} ^{\otimes p}}\!\!\!\!f_{\overrightarrow{\sigma}\!,\!\overrightarrow{\sigma}}\!\left(k_{\sigma_{p}}\!\left(-1\right)^{\!\overrightarrow{\sigma}}\!\right),\label{eq:SPA-asymptotics}
\end{equation}
where $\left(-1\right)^{\!\overrightarrow{\sigma}}=\left(\left(-1\right)^{\sigma_{1}},\ldots,\left(-1\right)^{\sigma_{p}}\right).$
Substituting the specific expressions for the functions $f_{\overrightarrow{\tau}\!,\!\overrightarrow{\sigma}}$
that satisfy Eq.~(\ref{eq:Correlation-Moments-Fully-Integral}),
we then find that
\begin{equation}
{\rm Tr}\!\left[\left\{ \left(C_{A_{i}}\right)^{n}+\left(\mathbb{I}-C_{A_{i}}\right)^{n}-\mathbb{I}\right\} ^{s}\right]\sim\ell_{i}\negthinspace\int_{k_{-}}^{k_{+}}\frac{dk}{2\pi}\left\{ \left({\cal T}\!\left(k\right)\right)^{n}+\left({\cal R}\!\left(k\right)\right)^{n}-1\right\} ^{s}\label{eq:SPA-asymptotics-explicit}
\end{equation}
for all positive integers $s$. A similar treatment was applied to
compute the leading term in $\Delta\ell_{i}$ of moments of $C_{A}$,
producing the same result as in Eq.~(\ref{eq:SPA-asymptotics-explicit})
up to replacing $C_{A_{i}}$ with $C_{A}$ and $\ell_{i}$ with $\Delta\ell_{{\scriptscriptstyle L}}+\Delta\ell_{{\scriptscriptstyle R}}$.

Summing up these contributions in the series expansion of Eq.~(\ref{eq:Renyi-from-correlations})
yields the extensive terms for the R\'enyi entropies of $A_{{\scriptscriptstyle L}}$,
$A_{{\scriptscriptstyle R}}$ and $A$,
\begin{align}
S_{A_{i}}^{\left(n\right)} & \sim\frac{\ell_{i}}{1-n}\int_{k_{-}}^{k_{+}}\frac{dk}{2\pi}\ln\left[\left({\cal T}\!\left(k\right)\right)^{n}+\left({\cal R}\!\left(k\right)\right)^{n}\right],\nonumber \\
S_{A}^{\left(n\right)} & \sim\frac{\Delta\ell_{{\scriptscriptstyle L}}+\Delta\ell_{{\scriptscriptstyle R}}}{1-n}\int_{k_{-}}^{k_{+}}\frac{dk}{2\pi}\ln\left[\left({\cal T}\!\left(k\right)\right)^{n}+\left({\cal R}\!\left(k\right)\right)^{n}\right].\label{eq:Renyi-entropies-asymptotics}
\end{align}
The asymptotics in Eq.~(\ref{eq:Renyi-entropies-asymptotics}) directly
lead to Eq.~(\ref{eq:Renyi_MI_volume_law}), while Eqs.~(\ref{eq:MI_volume_law})
and (\ref{eq:CI_volume_law}) are obtained by taking the limit $n\to1$
and substituting the resulting von Neumann entropies into the definitions
in Eqs.~(\ref{eq:MI-definition}) and (\ref{eq:CI-definition}).
We stress that the universal dependence of the R\'enyi entropies
on the scattering probabilities results from the fact that, at sites
$m$ lying outside the scattering region, the wavefunctions $u_{m}\!\left(k\right)$
in Eq.~(\ref{eq:Unnormalized-wavefunctions}) are written only in
terms of plane waves and scattering amplitudes.

\subsection{Fermionic negativity\label{subsec:Derivation_of_negativity}}

Here we outline the principal steps in the derivation of Eq.~(\ref{eq:Negativity_volume_law}),
the asymptotic formula for the fermionic negativity; a more detailed
account of the computation appears in Appendix~\ref{sec:Calculation-of-negativity}.
The derivation relies on the fact that the negativity ${\cal E}$
can be obtained as the analytic continuation of the R\'enyi negativities
${\cal E}_{n}=\ln{\rm Tr}\!\left[\left(\left(\widetilde{\rho}_{{\scriptscriptstyle A}}\right)^{\dagger}\widetilde{\rho}_{{\scriptscriptstyle A}}\right)^{n/2}\right]$
at $n=1$, where ${\cal E}_{n}$ are evaluated at even values of $n$~\citep{PhysRevB.95.165101}.
In analogy to the R\'enyi entropies, the R\'enyi negativities ${\cal E}_{n}$
can be expressed as functions of $C_{A}$ and of a transformed two-point
correlation matrix restricted to $A$~\citep{PhysRevB.95.165101,Shapourian_2019,PhysRevB.97.165123}.
As shown in Appendix~\ref{sec:Calculation-of-negativity}, this expression
for ${\cal E}_{n}$ leads to the following series expansion:
\begin{equation}
{\cal E}_{n}=\sum_{s=1}^{\infty}\frac{\left(-1\right)^{s+1}}{s}{\rm Tr}\!\left[\left\{ \prod_{\gamma=-\frac{n-1}{2}}^{\frac{n-1}{2}}\left(\mathbb{I}-C_{\gamma}\right)-\mathbb{I}\right\} ^{\!s}\,\right].\label{eq:Renyi_negativity_series_expansion}
\end{equation}
Here each $C_{\gamma}$ is a transformed version of $C_{A}$, given
by
\begin{equation}
C_{\gamma}=\left(\begin{array}{cc}
\left(1-e^{\frac{2\pi i\gamma}{n}}\right)\mathbb{I}_{\ell_{L}} & 0\\
0 & \left(1+e^{\frac{-2\pi i\gamma}{n}}\right)\mathbb{I}_{\ell_{R}}
\end{array}\right)C_{A},
\end{equation}
where the entries of $C_{A}$ are ordered such that the first $\ell_{{\scriptscriptstyle L}}$
indices correspond to sites in $A_{{\scriptscriptstyle L}}$, and
the next $\ell_{{\scriptscriptstyle R}}$ correspond to sites in $A_{{\scriptscriptstyle R}}$.

Eq.~(\ref{eq:Renyi_negativity_series_expansion}) reduces the calculation
of the R\'enyi negativities to that of terms of the form ${\rm Tr}\!\left[C_{\gamma_{1}}C_{\gamma_{2}}\ldots C_{\gamma_{p}}\right]$,
which, by using Eq.~(\ref{eq:Correlation_function_integral}), may
also be written as
\begin{equation}
{\rm Tr}\!\left[C_{\gamma_{1}}\ldots C_{\gamma_{p}}\right]=\int\!\!\frac{d^{p}k}{\left(2\pi\right)^{p}}\prod_{j=1}^{p}\!\left[\left(1-e^{\frac{2\pi i\gamma_{j}}{n}}\right)\!\!\sum_{m\in A_{L}}\!u_{m}\!\left(k_{j-1}\right)u_{m}^{*}\!\left(k_{j}\right)+\left(1+e^{\frac{-2\pi i\gamma_{j}}{n}}\right)\!\!\sum_{m\in A_{R}}\!u_{m}\!\left(k_{j-1}\right)u_{m}^{*}\!\left(k_{j}\right)\right],\label{eq:Renyi-negativity-integral-decomposition}
\end{equation}
where the integral is computed over the domain $\left[-k_{{\scriptscriptstyle F,R}},k_{{\scriptscriptstyle F,L}}\right]^{p}$.
The remainder of the calculation is similar in spirit to that of the
R\'enyi entropies: the explicit forms of the wavefunctions from Eq.~(\ref{eq:Unnormalized-wavefunctions})
are substituted into Eq.~(\ref{eq:Renyi-negativity-integral-decomposition});
Eq.~(\ref{eq:Renyi-negativity-integral-decomposition}) is rewritten
as a $2p$-dimensional integral using $p$ fictitious variables; and
finally, this $2p$-dimensional integral is estimated through a stationary
phase approximation (see Appendix~\ref{sec:Calculation-of-negativity}).
This process eventually leads to the following result for every positive
integer $s$:
\begin{align}
{\rm Tr}\!\left[\left\{ \prod_{\gamma=-\frac{n-1}{2}}^{\frac{n-1}{2}}\left(\mathbb{I}-C_{\gamma}\right)-\mathbb{I}\right\} ^{\!s}\,\right] & \sim\ell_{{\rm mirror}}\int_{k_{-}}^{k_{+}}\frac{dk}{2\pi}\left\{ \left[{\cal T}^{n/2}+{\cal R}^{n/2}\right]^{2}-1\right\} ^{\!s}+\left(\Delta\ell_{{\scriptscriptstyle L}}+\Delta\ell_{{\scriptscriptstyle R}}\right)\int_{k_{-}}^{k_{+}}\frac{dk}{2\pi}\left\{ {\cal T}^{n}+{\cal R}^{n}-1\right\} ^{s}.\label{eq:Renyi-negativity-expansion-terms-asymptotics}
\end{align}
Upon summation of the series in Eq.~(\ref{eq:Renyi_negativity_series_expansion}),
we find that the R\'enyi negativities are given by
\begin{align}
{\cal E}_{n} & \sim\ell_{{\rm mirror}}\int_{k_{-}}^{k_{+}}\frac{dk}{\pi}\ln\!\left[{\cal T}^{n/2}+{\cal R}^{n/2}\right]+\left(\Delta\ell_{{\scriptscriptstyle L}}+\Delta\ell_{{\scriptscriptstyle R}}\right)\int_{k_{-}}^{k_{+}}\frac{dk}{2\pi}\ln\!\left[{\cal T}^{n}+{\cal R}^{n}\right],\label{eq:Renyi-negativities-asymptotics}
\end{align}
 and the exact asymptotics of the negativity in Eq.~(\ref{eq:Negativity_volume_law})
is obtained once the limit $n\to1$ is finally taken.

\section{Discussion and outlook\label{sec:Discussion-and-outlook}}

In this work we derived the exact asymptotics of correlation measures
for a nonequilibrium steady state of 1D noninteracting fermions. We
have shown that this state hosts extensive long-range entanglement
between subsystems that are on opposite sides of a current-conserving
scatterer, provided that their distances from it are similar. The
volume-law terms of these measures stem from the extensive number
of single-particle wavepackets that originate in the high-chemical-potential
reservoir, which are split by the scatterer so that they are in a
superposition of being found in either one of the mirroring subsystems.
The correlation measures thus exhibit a simple and universal dependence
on scattering probabilities, allowing to clearly read off the necessary
and sufficient conditions for the generation of this strong long-range
entanglement. Apart from the requirement that the scatterer be non-trivial,
the essential ingredients are the absence of decoherence mechanisms,
along with the extensive excess of particles emerging from one of
the reservoirs.

We therefore expect the main features of our results to hold in a
wide class of nonequilibrium steady states, including those of integrable
interacting systems~\citep{10.21468/SciPostPhys.4.3.017,Alba_2019,PhysRevLett.96.216802},
as well as when the reservoirs are at finite temperatures, and when
the scatterer induces particle gain and loss \citep{10.21468/SciPostPhys.12.1.011}.
Similar features should also appear in the dynamics following a quench
where two decoupled half-infinite chains are prepared with unequal
fillings, and the scatterer is suddenly introduced~\citep{Bertini_2018,Eisler_2012,10.21468/SciPostPhys.8.3.036}.
All of these scenarios offer intriguing prospects for future studies.
It would also be interesting to study the interplay of the effects
uncovered by this work with decoherence, which could arise due to
an integrability-breaking impurity or when the system is coupled to
Lindblad baths~\citep{PhysRevB.98.235128,Bastianello_2019,PhysRevResearch.2.043052,PhysRevB.102.205131},
or the interplay of the same effects with the addition of quadratic
pairing terms \citep{Kitaev_2001}, which break charge conservation,
to the Hamiltonian of Eq.~(\ref{eq:Model-Hamiltonian}).

Realizations of such models with, e.g., ultracold atoms~\citep{Husmann1498}
should allow to experimentally extract entanglement measures~\citep{PhysRevLett.109.020504,PhysRevLett.109.020505,Islam2015,PhysRevLett.120.050406,PhysRevA.99.062309,doi:10.1126/science.aau4963}.
In this context we highlight our results in Eqs.~(\ref{eq:Renyi-entropies-asymptotics})
and (\ref{eq:Renyi-negativities-asymptotics}) for the R\'enyi versions
of these measures, which are generally more amenable to efficient
measurement than their von Neumann counterparts.

Replacing the scattering region with a disordered region~\citep{PhysRevX.9.021007,PhysRevLett.123.110601},
the signatures of localization and resonances on long-range entanglement
properties could also be a fruitful subject of research. Finally,
another possible future direction is the study of symmetry-resolution~\citep{PhysRevLett.120.200602,PhysRevA.98.032302,PhysRevB.100.235146,Bonsignori_2019,Fraenkel_2020,10.21468/SciPostPhys.8.3.046,Capizzi_2020,PhysRevB.101.235169,PhysRevB.102.014455,10.21468/SciPostPhys.10.3.054,Zhao2021}
of the different entanglement measures analyzed here.
\begin{acknowledgments}
We thank P.~Calabrese, V.~Eisler, and E.~Sela for fruitful discussions.
Our work was supported by the Israel Science Foundation (ISF) and
the Directorate for Defense Research and Development (DDR\&D) grant
No.~3427/21, and by the US-Israel Binational Science Foundation (BSF)
Grant No.~2020072. S.F.~is supported by the Azrieli Foundation Fellows
program.
\end{acknowledgments}

\appendix

\section{Two-point correlations\label{sec:Two-point-correlations}}

Here we derive Eq.~(\ref{eq:Correlation_function_integral}), which
is the general expression for the two-point correlation function $\left\langle c_{j}^{\dagger}c_{m}\right\rangle $
for sites outside the scattering region, $\left|j\right|,\left|m\right|>m_{0}$.
We consider a long chain with $N\gg1$ sites, where the small scattering
region is located at its center; in the end we will take the thermodynamic
limit $N\to\infty$.

An annihilation operator $c_{m}$ may be expanded in terms of annihilation
operators corresponding to the single-particle energy eigenstates.
We include only extended scattering states in this expansion, neglecting
the contribution of localized bound states, since the amplitude of
a bound state wavefunction at any site outside the scattering region
decays exponentially with the distance of that site from the scatterer.
More concretely, we associate an annihilation operator $c_{{\scriptscriptstyle k,L}}$
($c_{{\scriptscriptstyle k,R}}$) with the scattering state of a particle
originating in the left (right) reservoir with momentum $k>0$ ($k<0$).
Then, $c_{m}$ may be written as follows:
\begin{equation}
c_{m}=\frac{1}{\sqrt{N}}\left[\sum_{-\pi<k<0}u_{m}\!\left(k\right)c_{{\scriptscriptstyle k,R}}+\sum_{0<k<\pi}u_{m}\!\left(k\right)c_{{\scriptscriptstyle k,L}}\right],\label{eq:Scattering_state_expansion}
\end{equation}
where the wavefunctions $u_{m}\!\left(k\right)$ are given in Eq.~(\ref{eq:Unnormalized-wavefunctions}).
In the language of scattering state creation operators, the nonequilibrium
steady state analyzed in this work is given by
\begin{equation}
|{\rm NESS}\rangle=\left(\prod_{-k_{F,R}<k<0}c_{{\scriptscriptstyle k,R}}^{\dagger}\right)\left(\prod_{0<k<k_{F,L}}c_{{\scriptscriptstyle k,L}}^{\dagger}\right)|{\rm vac}\rangle,
\end{equation}
with $|{\rm vac}\rangle$ being the vacuum state. Substituting Eq.~(\ref{eq:Scattering_state_expansion})
into the definition of the two-point correlation function, we find
that, in the thermodynamic limit $N\to\infty$, the correlation function
approaches the integral expression of Eq.~(\ref{eq:Correlation_function_integral}).

As explained in Sec.~\ref{sec:Analytical-method}, the different
correlation measures discussed in this work can all be expressed as
functions of two-point correlation matrices restricted to the subsystems
of interest. That is, we generally consider the terms given by Eq.~(\ref{eq:Correlation_function_integral})
only for sites $j,m\in A$, and correlation measures can scale at
most as ${\cal O}\!\left(\ell_{{\scriptscriptstyle L}}+\ell_{{\scriptscriptstyle R}}\right)$,
given the dimensions of the correlation matrices. This, in turn, implies
that in the limit $d_{i}/\ell_{i}\to\infty$ (with $d_{{\scriptscriptstyle L}}-d_{{\scriptscriptstyle R}}$
kept fixed), calculations of correlation measures can be simplified
by first neglecting certain terms in the expressions for the matrix
elements $\left\langle c_{j}^{\dagger}c_{m}\right\rangle $, and only
then calculating the appropriate functions of the correlation matrices.

In particular, using Eq.~(\ref{eq:Unnormalized-wavefunctions}) we
observe that in Eq.~(\ref{eq:Correlation_function_integral}) the
correlation function is a sum of integrals, where in each integral
the integrand is a product of a function of $k$ that is independent
of $j,m$ and an exponent of the form $\exp\left[i\alpha_{j,m}k\right]$,
with $\alpha_{j,m}\in\left\{ \pm\left(j\pm m\right)\right\} $. Then,
the Riemann-Lebesgue lemma leads to the conclusion that when $\left|\alpha_{j,m}\right|\gg\ell_{{\scriptscriptstyle L}},\ell_{{\scriptscriptstyle R}}$,
the contribution of the integral is negligible and may be omitted.
This entails that when $d_{{\scriptscriptstyle L}},d_{{\scriptscriptstyle R}}\gg\ell_{{\scriptscriptstyle L}},\ell_{{\scriptscriptstyle R}}$
we may use the following approximations for the correlation matrix
elements:
\begin{equation}
\left\langle c_{j}^{\dagger}c_{m}\right\rangle \approx\begin{cases}
\int_{-k_{F,R}}^{k_{F,R}}\frac{dk}{2\pi}e^{-i\left(j-m\right)k}+\int_{k_{F,R}}^{k_{F,L}}\frac{dk}{2\pi}{\cal T}\!\left(k\right)e^{-i\left(j-m\right)k} & j,m\in A_{{\scriptscriptstyle R}},\\
\int_{-k_{F,L}}^{k_{F,L}}\frac{dk}{2\pi}e^{-i\left(j-m\right)k}+\int_{k_{F,L}}^{k_{F,R}}\frac{dk}{2\pi}{\cal T}\!\left(k\right)e^{i\left(j-m\right)k} & j,m\in A_{{\scriptscriptstyle L}},\\
\int_{k_{F,R}}^{k_{F,L}}\frac{dk}{2\pi}t_{{\scriptscriptstyle L}}^{*}\!\left(k\right)r_{{\scriptscriptstyle L}}\!\left(k\right)e^{-i\left(j+m\right)k} & m\in A_{{\scriptscriptstyle L}}\text{ and }j\in A_{{\scriptscriptstyle R}},\\
\int_{k_{F,L}}^{k_{F,R}}\frac{dk}{2\pi}t_{{\scriptscriptstyle R}}^{*}\!\left(k\right)r_{{\scriptscriptstyle R}}\!\left(k\right)e^{i\left(j+m\right)k} & j\in A_{{\scriptscriptstyle L}}\text{ and }m\in A_{{\scriptscriptstyle R}}.
\end{cases}\label{eq:Correlation-function-large-distance}
\end{equation}
Our analytical results were all derived based on the full expression
for $\left\langle c_{j}^{\dagger}c_{m}\right\rangle $ in Eq.~(\ref{eq:Correlation_function_integral}).
As they indicated that the volume-law terms of the different correlation
measures depend on $d_{{\scriptscriptstyle L}}$ and $d_{{\scriptscriptstyle R}}$
only through $d_{{\scriptscriptstyle L}}-d_{{\scriptscriptstyle R}}$,
they were compared in Figs.~\ref{fig:Num-Comparison-Symmetric}--\ref{fig:MI-Varying-Distance}
to numerical results that were computed in the limit $d_{i}/\ell_{i}\to\infty$,
based on the approximated correlation function in Eq.~(\ref{eq:Correlation-function-large-distance}).
A comparison to a numerical calculation that relies on the full expression
for the correlation function is provided in Appendix~\ref{sec:Additional-numerical-tests}.

\section{Calculation of the R\'enyi entropies\label{sec:Calculation-of-Renyi-entropies}}

In this appendix we describe the analytical method used for the computation
of the R\'enyi entropies $S_{X}^{\left(n\right)}=\frac{1}{1-n}\ln{\rm Tr}\left[\left(\rho_{{\scriptscriptstyle X}}\right)^{n}\right]$
for the subsystems $X=A_{{\scriptscriptstyle L}},A_{{\scriptscriptstyle R}},A$.
The final results are given in Eq.~(\ref{eq:Renyi-entropies-asymptotics}).
The results for the R\'enyi entropies lead directly to the asymptotics
of the MI and CI (Eqs.~(\ref{eq:MI_volume_law}) and (\ref{eq:CI_volume_law})),
as explained in Subsec.~\ref{subsec:Derivation_of_MI_CI}.

In Subsec.~\ref{subsec:Derivation_of_MI_CI} we showed that the calculation
of R\'enyi entropies can be reduced to that of the moments ${\rm Tr}\!\left[\left(C_{X}\right)^{p}\right]$
for all positive integers $p$. We now derive the asymptotics of these
moments for the subsystems of interest, starting from their integral
expression in Eq.~(\ref{eq:Correlation-Moments}). The analysis is
based on the stationary phase approximation (SPA)~\citep{doi:10.1137/1.9780898719260},
and is inspired by the analytical methods of Refs.~\citep{PhysRevA.78.010306,10.21468/SciPostPhys.12.1.011}.

\subsubsection{Asymptotics of moments for the connected subsystems}

We first consider the case $X=A_{{\scriptscriptstyle R}}$. We begin
by introducing the notation ${\cal W}_{{\scriptscriptstyle R}}\!\left(x\right)=\frac{x}{\sin x}\exp\!\left[2i\left(m_{0}+d_{{\scriptscriptstyle R}}+\frac{1}{2}\right)x\right]$,
and observing that
\begin{equation}
\sum_{m=m_{0}+d_{R}+1}^{m_{0}+d_{R}+\ell_{R}}\exp\!\left[im\left(k_{j-1}-k_{j}\right)\right]=\frac{\ell_{{\scriptscriptstyle R}}}{2}{\cal W}_{{\scriptscriptstyle R}}\!\left(\frac{k_{j-1}-k_{j}}{2}\right)\underset{-1}{\overset{1}{\int}}d\xi\exp\!\left[i\frac{\ell_{{\scriptscriptstyle R}}}{2}\left(k_{j-1}-k_{j}\right)\left(\xi+1\right)\right].\label{eq:Elementary-sum-as-integral}
\end{equation}
For convenience, we define the notation $k_{a_{j}}=\left(-1\right)^{a_{j}}k_{j}$
for $a_{j}\in\left\{ 0,1\right\} $, as well as $\left(-1\right)^{\overrightarrow{a}}=\left(\left(-1\right)^{a_{1}},\ldots,\left(-1\right)^{a_{p}}\right)$
for $\overrightarrow{a}\in\left\{ 0,1\right\} ^{\otimes p}$. We use
Eqs.~(\ref{eq:Unnormalized-wavefunctions}) and (\ref{eq:Elementary-sum-as-integral})
to write
\begin{equation}
\sum_{m\in A_{R}}u_{m}\!\left(k_{j-1}\right)u_{m}^{*}\!\left(k_{j}\right)=\frac{\ell_{{\scriptscriptstyle R}}}{2}\sum_{a_{j-1},b_{j}=0,1}\Xi^{a_{j-1}b_{j}}\!\left(k_{a_{j-1}},k_{b_{j}}\right)\Theta\!\left(k_{a_{j-1}}\right)\Theta\!\left(k_{b_{j}}\right),\label{eq:Xi-divided-into-domains}
\end{equation}
where $\Theta\!\left(x\right)$ is the Heaviside step function, and
where we defined
\begin{align}
\Xi^{00}\!\left(k_{j-1},k_{j}\right) & =t_{{\scriptscriptstyle L}}\!\left(\left|k_{j-1}\right|\right)t_{{\scriptscriptstyle L}}^{*}\!\left(\left|k_{j}\right|\right){\cal W}_{{\scriptscriptstyle R}}\!\left(\frac{k_{j-1}-k_{j}}{2}\right)\int_{-1}^{1}d\xi e{}^{\frac{i}{2}\ell_{R}\left(k_{j-1}-k_{j}\right)\left(\xi+1\right)},\nonumber \\
\Xi^{11}\!\left(k_{j-1},k_{j}\right) & =\int_{-1}^{1}d\xi\left\{ {\cal W}_{{\scriptscriptstyle R}}\!\left(\frac{k_{j}-k_{j-1}}{2}\right)e{}^{\frac{i}{2}\ell_{R}\left(k_{j}-k_{j-1}\right)\left(\xi+1\right)}+r_{{\scriptscriptstyle R}}\!\left(\left|k_{j-1}\right|\right)r_{{\scriptscriptstyle R}}^{*}\!\left(\left|k_{j}\right|\right){\cal W}_{{\scriptscriptstyle R}}\!\left(\frac{k_{j-1}-k_{j}}{2}\right)e^{\frac{i}{2}\ell_{R}\left(k_{j-1}-k_{j}\right)\left(\xi+1\right)}\right\} \nonumber \\
 & +\int_{-1}^{1}d\xi\left\{ r_{{\scriptscriptstyle R}}^{*}\!\left(\left|k_{j}\right|\right){\cal W}_{{\scriptscriptstyle R}}\!\left(\frac{-k_{j-1}-k_{j}}{2}\right)e{}^{-\frac{i}{2}\ell_{R}\left(k_{j-1}+k_{j}\right)\left(\xi+1\right)}+r_{{\scriptscriptstyle R}}\!\left(\left|k_{j-1}\right|\right){\cal W}_{{\scriptscriptstyle R}}\!\left(\frac{k_{j-1}+k_{j}}{2}\right)e^{\frac{i}{2}\ell_{R}\left(k_{j-1}+k_{j}\right)\left(\xi+1\right)}\right\} ,\nonumber \\
\Xi^{01}\!\left(k_{j-1},k_{j}\right) & =\int_{-1}^{1}d\xi\,t_{{\scriptscriptstyle L}}\!\left(\left|k_{j-1}\right|\right)\left\{ {\cal W}_{{\scriptscriptstyle R}}\!\left(\frac{k_{j-1}+k_{j}}{2}\right)e{}^{\frac{i}{2}\ell_{R}\left(k_{j-1}+k_{j}\right)\left(\xi+1\right)}+r_{{\scriptscriptstyle R}}^{*}\!\left(\left|k_{j}\right|\right){\cal W}_{{\scriptscriptstyle R}}\!\left(\frac{k_{j-1}-k_{j}}{2}\right)e^{\frac{i}{2}\ell_{R}\left(k_{j-1}-k_{j}\right)\left(\xi+1\right)}\right\} ,\label{eq:Xi-domains-explicit}
\end{align}
and $\Xi^{10}\!\left(k_{j-1},k_{j}\right)=\Xi^{01}\!\left(k_{j},k_{j-1}\right)^{*}$.

When plugging Eq.~(\ref{eq:Xi-divided-into-domains}) into the expression
for ${\rm Tr}\!\left[\left(C_{A_{R}}\right)^{p}\right]$ in Eq.~(\ref{eq:Correlation-Moments}),
we will generally get a sum of $2p$-dimensional integrals, each of
the form
\begin{equation}
{\cal F}\!\left(\overrightarrow{\tau},\overrightarrow{\sigma}\right)=\left(\frac{\ell_{{\scriptscriptstyle R}}}{4\pi}\right)^{p}\int_{\left[-k_{F,R},k_{F,L}\right]^{p}}d^{p}k\int_{\left[-1,1\right]^{p}}d^{p}\xi\,f_{\overrightarrow{\tau}\!,\!\overrightarrow{\sigma}}\!\left(\overrightarrow{k}\right)\exp\left[i\frac{\ell_{{\scriptscriptstyle R}}}{2}\sum_{j=1}^{p}\left(k_{\tau_{j-1}}-k_{\sigma_{j}}\right)\left(\xi_{j}+1\right)\right],\label{eq: General-integral-expression}
\end{equation}
with $\overrightarrow{\tau},\overrightarrow{\sigma}\in\left\{ 0,1\right\} ^{\otimes p}$,
and where the function $f_{\overrightarrow{\tau}\!,\!\overrightarrow{\sigma}}\!\left(\overrightarrow{k}\right)$
is independent of $\ell_{{\scriptscriptstyle R}}$ and supported on
$\left[-k_{{\scriptscriptstyle F,R}},k_{{\scriptscriptstyle F,L}}\right]^{p}$.
We apply a change of variables
\begin{align}
\zeta_{1} & =\xi_{1},\nonumber \\
\zeta_{j} & =\xi_{j}-\xi_{j-1}\,\,\,\,\,(2\le j\le p),\label{eq:SPA-Change-of-variables}
\end{align}
and obtain
\begin{align}
{\cal F}\!\left(\overrightarrow{\tau},\overrightarrow{\sigma}\right) & =\left(\frac{\ell_{{\scriptscriptstyle R}}}{4\pi}\right)^{p}\int_{\left[-k_{F,R},k_{F,L}\right]^{p}}d^{p}k\int d^{p}\zeta\,f_{\overrightarrow{\tau}\!,\!\overrightarrow{\sigma}}\!\left(\overrightarrow{k}\right)\exp\left[i\frac{\ell_{{\scriptscriptstyle R}}}{2}\left\{ \sum_{j=1}^{p}\left(k_{\tau_{j-1}}-k_{\sigma_{j}}\right)+\sum_{l=1}^{p}\zeta_{l}\sum_{j=l}^{p}\left(k_{\tau_{j-1}}-k_{\sigma_{j}}\right)\right\} \right].\label{eq:General-integral-expression-variable-change}
\end{align}
These are the integrals to which we apply the SPA. The SPA allows
us to detect which integrals contribute to the leading-order terms
of ${\rm Tr}\left[\left(C_{A_{R}}\right)^{p}\right]$, and to compute
their exact contribution to the linear term in $\ell_{{\scriptscriptstyle R}}$.
From Eq.~(\ref{eq:General-integral-expression-variable-change})
it is evident that the answer to the question of whether ${\cal F}\!\left(\overrightarrow{\tau},\overrightarrow{\sigma}\right)$
has a leading-order contribution is determined by the values of $\overrightarrow{\tau}$
and $\overrightarrow{\sigma}$. We now illustrate the method by focusing
on two concrete cases for the choice of $\overrightarrow{\tau}$ and
$\overrightarrow{\sigma}$.

Assuming that $\tau_{j}=\sigma_{j}$ for every $j$, we find that
\begin{align}
{\cal F}\!\left(\overrightarrow{\sigma},\overrightarrow{\sigma}\right) & =\left(\frac{\ell_{{\scriptscriptstyle R}}}{4\pi}\right)^{p}\int_{-k_{F,R}}^{k_{F,L}}\!\!dk_{p}\int_{-1}^{1}d\zeta_{1}\int d^{p-1}k\,d^{p-1}\zeta\,f_{\overrightarrow{\sigma}\!,\!\overrightarrow{\sigma}}\!\left(\overrightarrow{k}\right)\exp\left[i\frac{\ell_{{\scriptscriptstyle R}}}{2}\sum_{l=2}^{p}\zeta_{l}\left(k_{\sigma_{l-1}}-k_{\sigma_{p}}\right)\right].
\end{align}
Applying the SPA to the innermost $\left(2p-2\right)$-dimensional
integral, with respect to the stationary point of the function $\frac{1}{2}\sum_{l=2}^{p}\zeta_{l}\left(k_{\sigma_{l-1}}-k_{\sigma_{p}}\right)$
(at which $\zeta_{l}=0$ and $k_{\sigma_{l-1}}=k_{\sigma_{p}}$ for
$l=2,\ldots,p$), we obtain~\citep{doi:10.1137/1.9780898719260}
\begin{align}
{\cal F}\!\left(\overrightarrow{\sigma},\overrightarrow{\sigma}\right) & \sim\left(\frac{\ell_{{\scriptscriptstyle R}}}{4\pi}\right)^{p}\int_{-k_{F,R}}^{k_{F,L}}\!\!dk_{p}\int_{-1}^{1}d\zeta_{1}\,f_{\overrightarrow{\sigma}\!,\!\overrightarrow{\sigma}}\!\left(k_{\sigma_{p}}\left(-1\right)^{\overrightarrow{\sigma}}\right)\left\{ \left(\frac{2\pi}{\ell_{{\scriptscriptstyle R}}}\right)^{p-1}\left|\det H\right|^{-1/2}\right\} =\frac{\ell_{{\scriptscriptstyle R}}}{2\pi}\int_{-k_{F,R}}^{k_{F,L}}\!\!dk_{p}\,f_{\overrightarrow{\sigma}\!,\!\overrightarrow{\sigma}}\!\left(k_{\sigma_{p}}\left(-1\right)^{\overrightarrow{\sigma}}\right),\label{eq:Leading-order-contributing-integrals}
\end{align}
where we used the fact that the Hessian $H$ at the stationary point
yields $\left|\det H\right|=\left(\frac{1}{2}\right)^{2p-2}$.

If, on the other hand, $\tau_{j}=\sigma_{j}$ for every $j\ge2$ but
$\tau_{1}\neq\sigma_{1}$, then
\begin{align}
{\cal F}\!\left(\overrightarrow{\tau},\overrightarrow{\sigma}\right) & =\left(\frac{\ell_{{\scriptscriptstyle R}}}{4\pi}\right)^{p}\int_{-k_{F,R}}^{k_{F,L}}\!\!dk_{1}\int_{-1}^{1}d\zeta_{1}\exp\left[i\ell_{{\scriptscriptstyle R}}k_{\tau_{1}}\left(\zeta_{1}+1\right)\right]\int d^{p-1}k\,d^{p-1}\zeta\,f_{\overrightarrow{\tau}\!,\!\overrightarrow{\sigma}}\!\left(\overrightarrow{k}\right)\exp\left[i\frac{\ell_{{\scriptscriptstyle R}}}{2}\sum_{l=2}^{p}\zeta_{l}\left(k_{\tau_{l-1}}-k_{\tau_{p}}\right)\right].
\end{align}
Applying the SPA to the innermost integral will again produce a factor
proportional to $\ell_{{\scriptscriptstyle R}}^{-p+1}$, yielding
\begin{equation}
{\cal F}\!\left(\overrightarrow{\tau},\overrightarrow{\sigma}\right)\sim\frac{\ell_{{\scriptscriptstyle R}}}{4\pi}\int_{-k_{F,R}}^{k_{F,L}}\!\!dk_{1}\int_{-1}^{1}d\zeta_{1}\,f_{\overrightarrow{\tau}\!,\!\overrightarrow{\sigma}}\!\left(k_{\tau_{1}}\left(-1\right)^{\overrightarrow{\tau}}\right)\exp\left[i\ell_{{\scriptscriptstyle R}}k_{\tau_{1}}\left(\zeta_{1}+1\right)\right],
\end{equation}
only now the remaining phase factor $\exp\left[i\ell_{{\scriptscriptstyle R}}k_{\tau_{1}}\left(\zeta_{1}+1\right)\right]$
will eliminate the extensive contribution, such that ${\cal F}\!\left(\overrightarrow{\tau},\overrightarrow{\sigma}\right)$
can have a contribution that is, at most, constant in $\ell_{{\scriptscriptstyle R}}$.

From these two examples, it is straightforward to infer the more general
rule that an integral ${\cal F}\!\left(\overrightarrow{\tau},\overrightarrow{\sigma}\right)$
can contribute to ${\rm Tr}\!\left[\left(C_{A_{R}}\right)^{p}\right]$
beyond the constant-in-$\ell_{{\scriptscriptstyle R}}$ order only
if $\overrightarrow{\sigma}=\overrightarrow{\tau}$. Furthermore,
if indeed $\overrightarrow{\sigma}=\overrightarrow{\tau}$, Eq.~(\ref{eq:Leading-order-contributing-integrals})
indicates the contribution of ${\cal F}\!\left(\overrightarrow{\tau},\overrightarrow{\sigma}\right)$
to the linear-in-$\ell_{{\scriptscriptstyle R}}$ term of ${\rm Tr}\!\left[\left(C_{A_{R}}\right)^{p}\right]$.

Let us now apply this general conclusion to our problem. Substituting
Eq.~(\ref{eq:Xi-divided-into-domains}) into Eq.~(\ref{eq:Correlation-Moments}),
we obtain
\begin{equation}
{\rm Tr}\!\left[\left(C_{A_{R}}\right)^{p}\right]=\left(\frac{\ell_{{\scriptscriptstyle R}}}{4\pi}\right)^{p}\int_{\left[-k_{F,R},k_{F,L}\right]^{p}}d^{p}k\sum_{\overrightarrow{a}\in\left\{ 0,1\right\} ^{\otimes p}}\prod_{j=1}^{p}\left[\Xi^{a_{j-1}a_{j}}\!\left(k_{a_{j-1}},k_{a_{j}}\right)\Theta\!\left(k_{a_{j}}\right)\right].\label{eq:Trace-moment-before-SPA}
\end{equation}
Eq.~(\ref{eq:Leading-order-contributing-integrals}) tells us that
the focus on leading-order terms confines ${\cal F}\!\left(\overrightarrow{\sigma},\overrightarrow{\sigma}\right)$
to an integration subdomain where $k_{\sigma_{j}}=k_{\sigma_{p}}$
for all $1\le j\le p$; this implies that, for the purpose of calculating
the leading-order asymptotics of ${\rm Tr}\!\left[\left(C_{A_{R}}\right)^{p}\right]$,
some of the terms in the full expressions for $\Xi^{a_{j-1}a_{j}}\!\left(k_{a_{j-1}},k_{a_{j}}\right)$
in Eq.~(\ref{eq:Xi-domains-explicit}) may be \emph{a priori} discarded,
given that for them the $k_{\sigma_{j-1}}=k_{\sigma_{j}}$ requirement
is satisfied only when $k_{\sigma_{j-1}}=k_{\sigma_{j}}=0$. Namely,
we may replace
\begin{align}
\Xi^{11}\!\left(k_{j-1},k_{j}\right) & \longrightarrow\int_{-1}^{1}d\xi\left\{ {\cal W}_{{\scriptscriptstyle R}}\!\left(\frac{k_{j}-k_{j-1}}{2}\right)e{}^{\frac{i}{2}\ell_{R}\left(k_{j}-k_{j-1}\right)\left(\xi+1\right)}+r_{{\scriptscriptstyle R}}\!\left(\left|k_{j-1}\right|\right)r_{{\scriptscriptstyle R}}^{*}\!\left(\left|k_{j}\right|\right){\cal W}_{{\scriptscriptstyle R}}\!\left(\frac{k_{j-1}-k_{j}}{2}\right)e^{\frac{i}{2}\ell_{R}\left(k_{j-1}-k_{j}\right)\left(\xi+1\right)}\right\} ,\nonumber \\
\Xi^{01}\!\left(k_{j-1},k_{j}\right) & \longrightarrow\int_{-1}^{1}d\xi\,t_{{\scriptscriptstyle L}}\!\left(\left|k_{j-1}\right|\right)r_{{\scriptscriptstyle R}}^{*}\!\left(\left|k_{j}\right|\right){\cal W}_{{\scriptscriptstyle R}}\!\left(\frac{k_{j-1}-k_{j}}{2}\right)e^{\frac{i}{2}\ell_{R}\left(k_{j-1}-k_{j}\right)\left(\xi+1\right)},\label{eq:Xi-domains-SPA}
\end{align}
and again $\Xi^{10}\!\left(k_{j-1},k_{j}\right)=\Xi^{01}\!\left(k_{j},k_{j-1}\right)^{*}$.
Note that in the first summand appearing in the expression for $\Xi^{11}\!\left(k_{j-1},k_{j}\right)$,
the term in the exponent has an opposite sign compared to all other
integrals (including $\Xi^{00}\!\left(k_{j-1},k_{j}\right)$). Since
we have established that expressions of the form ${\cal F}\left(\overrightarrow{\tau},\overrightarrow{\sigma}\right)$
in Eq.~(\ref{eq: General-integral-expression}) contribute to the
leading order only when $\overrightarrow{\sigma}=\overrightarrow{\tau}$,
a leading-order contribution to Eq.~(\ref{eq:Trace-moment-before-SPA})
will arise from this integral only for $\overrightarrow{a}=1^{\otimes p}$,
meaning that we can write 
\begin{align}
{\rm Tr}\left[\left(C_{A_{R}}\right)^{p}\right] & \sim\left(\frac{\ell_{{\scriptscriptstyle R}}}{4\pi}\right)^{p}\int_{\left[-k_{F,R},k_{F,L}\right]^{p}}\!\!\!d^{p}k\int_{\left[-1,1\right]^{p}}\!\!\!d^{p}\xi\prod_{j=1}^{p}\left[\Theta\!\left(-k_{j}\right){\cal W}_{{\scriptscriptstyle R}}\!\left(\frac{k_{j-1}-k_{j}}{2}\right)\exp\left[\frac{i\ell_{{\scriptscriptstyle R}}}{2}\left(k_{j-1}-k_{j}\right)\left(\xi_{j}+1\right)\right]\right]\nonumber \\
 & +\left(\frac{\ell_{{\scriptscriptstyle R}}}{4\pi}\right)^{p}\int_{\left[-k_{F,R},k_{F,L}\right]^{p}}\!\!\!d^{p}k\int_{\left[-1,1\right]^{p}}\!\!\!d^{p}\xi\sum_{\overrightarrow{a}\in\left\{ 0,1\right\} ^{\otimes p}}\prod_{j=1}^{p}\left[\Theta\!\left(k_{a_{j}}\right){\cal W}_{{\scriptscriptstyle R}}\!\left(\frac{k_{a_{j-1}}-k_{a_{j}}}{2}\right)\exp\left[\frac{i\ell_{{\scriptscriptstyle R}}}{2}\left(k_{a_{j-1}}-k_{a_{j}}\right)\left(\xi_{j}+1\right)\right]\right]\nonumber \\
 & \times\frac{1+\left(-1\right)^{a_{j}}\!\left[{\cal T}\!\left(k_{a_{j}}\right)-{\cal R}\!\left(k_{a_{j}}\right)\right]}{2}.
\end{align}
Applying the SPA as explained above while using the fact that ${\cal W}_{{\scriptscriptstyle R}}\!\left(0\right)=1$,
we thus have
\begin{equation}
{\rm Tr}\!\left[\left(C_{A_{R}}\right)^{p}\right]\sim\ell_{{\scriptscriptstyle R}}\begin{cases}
\frac{k_{{\scriptscriptstyle F,R}}}{\pi}+\int_{k_{F,R}}^{k_{F,L}}\frac{dk}{2\pi}\left({\cal T}\!\left(k\right)\right)^{p} & k_{{\scriptscriptstyle F,L}}>k_{{\scriptscriptstyle F,R}},\\
\frac{k_{{\scriptscriptstyle F,L}}+k_{{\scriptscriptstyle F,R}}}{2\pi}+\int_{k_{F,L}}^{k_{F,R}}\frac{dk}{2\pi}\left({\cal R}\!\left(k\right)\right)^{p} & k_{{\scriptscriptstyle F,L}}<k_{{\scriptscriptstyle F,R}}.
\end{cases}\label{eq:Moment-asymptotics-A_R}
\end{equation}
The derivation for the case $X=A_{{\scriptscriptstyle L}}$ is equivalent,
yielding
\begin{equation}
{\rm Tr}\!\left[\left(C_{A_{L}}\right)^{p}\right]\sim\ell_{{\scriptscriptstyle L}}\begin{cases}
\frac{k_{{\scriptscriptstyle F,L}}+k_{{\scriptscriptstyle F,R}}}{2\pi}+\int_{k_{F,R}}^{k_{F,L}}\frac{dk}{2\pi}\left({\cal R}\!\left(k\right)\right)^{p} & k_{{\scriptscriptstyle F,L}}>k_{{\scriptscriptstyle F,R}},\\
\frac{k_{{\scriptscriptstyle F,L}}}{\pi}+\int_{k_{F,L}}^{k_{F,R}}\frac{dk}{2\pi}\left({\cal T}\!\left(k\right)\right)^{p} & k_{{\scriptscriptstyle F,L}}<k_{{\scriptscriptstyle F,R}}.
\end{cases}\label{eq:Moment-asymptotics-A_L}
\end{equation}

\subsubsection{Asymptotics of moments for the disjoint subsystem}

We now consider the case $X=A$. In the summation over sites $m$
in $\sum_{m\in A}u_{m}\left(k_{j-1}\right)u_{m}^{*}\left(k_{j}\right)$
(the sum that appears in Eq.~(\ref{eq:Correlation-Moments})) we
will separate mirroring sites from sites which are not mirrored. For
concreteness, we assume that $d_{{\scriptscriptstyle L}}<d_{{\scriptscriptstyle R}}<d_{{\scriptscriptstyle L}}+\ell_{{\scriptscriptstyle L}}<d_{{\scriptscriptstyle R}}+\ell_{{\scriptscriptstyle R}}$,
where the subsequent generalization is straightforward. We then have
\begin{align}
\sum_{m\in A}u_{m}\!\left(k_{j-1}\right)u_{m}^{*}\!\left(k_{j}\right) & =\sum_{m=m_{0}+d_{L}+1}^{m_{0}+d_{R}}u_{-m}\!\left(k_{j-1}\right)u_{-m}^{*}\!\left(k_{j}\right)+\sum_{m=m_{0}+d_{{\scriptscriptstyle L}}+\ell_{{\scriptscriptstyle L}}+1}^{m_{0}+d_{R}+\ell_{{\scriptscriptstyle R}}}u_{m}\!\left(k_{j-1}\right)u_{m}^{*}\!\left(k_{j}\right)\nonumber \\
 & +\sum_{m=m_{0}+d_{R}+1}^{m_{0}+d_{L}+\ell_{L}}\left[u_{m}\!\left(k_{j-1}\right)u_{m}^{*}\!\left(k_{j}\right)+u_{-m}\!\left(k_{j-1}\right)u_{-m}^{*}\!\left(k_{j}\right)\right].\label{eq:Xi-disjoint-subsystem}
\end{align}
We define the function ${\cal W}_{{\scriptscriptstyle L}}\!\left(x\right)=\frac{x}{\sin x}\exp\left[2i\left(m_{0}+d_{{\scriptscriptstyle L}}+\frac{1}{2}\right)x\right]$.
Sums of exponents appearing in Eq.~(\ref{eq:Xi-disjoint-subsystem})
can be written as integrals:
\begin{align}
\sum_{m=m_{0}+d_{L}+1}^{m_{0}+d_{R}}\exp\left[im\left(k_{j-1}-k_{j}\right)\right] & =\frac{\Delta\ell_{{\scriptscriptstyle L}}}{2}{\cal W}_{{\scriptscriptstyle L}}\!\left(\frac{k_{j-1}-k_{j}}{2}\right)\underset{-1}{\overset{1}{\int}}d\xi\exp\left\{ i\left(k_{j-1}-k_{j}\right)\left[\frac{\Delta\ell_{{\scriptscriptstyle L}}}{2}\left(\xi+1\right)\right]\right\} ,\nonumber \\
\sum_{m=m_{0}+d_{R}+1}^{m_{0}+d_{L}+\ell_{L}}\exp\left[im\left(k_{j-1}-k_{j}\right)\right] & =\frac{\ell_{{\rm mirror}}}{2}{\cal W}_{{\scriptscriptstyle L}}\!\left(\frac{k_{j-1}-k_{j}}{2}\right)\underset{-1}{\overset{1}{\int}}d\xi\exp\left\{ i\left(k_{j-1}-k_{j}\right)\left[\frac{\ell_{{\rm mirror}}}{2}\left(\xi+1\right)+\Delta\ell_{{\scriptscriptstyle L}}\right]\right\} ,\nonumber \\
\sum_{m=m_{0}+d_{{\scriptscriptstyle L}}+\ell_{{\scriptscriptstyle L}}+1}^{m_{0}+d_{R}+\ell_{{\scriptscriptstyle R}}}\exp\left[im\left(k_{j-1}-k_{j}\right)\right] & =\frac{\Delta\ell_{{\scriptscriptstyle R}}}{2}{\cal W}_{{\scriptscriptstyle L}}\!\left(\frac{k_{j-1}-k_{j}}{2}\right)\underset{-1}{\overset{1}{\int}}d\xi\exp\left\{ i\left(k_{j-1}-k_{j}\right)\left[\frac{\Delta\ell_{{\scriptscriptstyle R}}}{2}\left(\xi+1\right)+\ell_{{\scriptscriptstyle L}}\right]\right\} .
\end{align}
The substitution of Eq.~(\ref{eq:Xi-disjoint-subsystem}) into the
integral expression for ${\rm Tr}\!\left[\left(C_{A}\right)^{p}\right]$
in Eq.~(\ref{eq:Correlation-Moments}) will then yield a sum of integrals
of the form
\begin{equation}
{\cal F}\!\left(\overrightarrow{\tau},\overrightarrow{\sigma},\overrightarrow{{\cal A}}\right)=\left[\prod_{j=1}^{p}\frac{{\cal A}_{j}}{2}\right]\int_{\left[-k_{F,R},k_{F,L}\right]^{p}}\frac{d^{p}k}{\left(2\pi\right)^{p}}\int_{\left[-1,1\right]^{p}}d^{p}\xi\,f\!\left(\overrightarrow{k}\right)\exp\left\{ i\sum_{j=1}^{p}\left(k_{\tau_{j-1}}-k_{\sigma_{j}}\right)\left[\frac{{\cal A}_{j}}{2}\left(\xi_{j}+1\right)+{\cal B}_{j}\right]\right\} ,
\end{equation}
where $\left({\cal A}_{j},{\cal B}_{j}\right)\in\left\{ \left(\Delta\ell_{{\scriptscriptstyle L}},0\right),\left(\ell_{{\rm mirror}},\Delta\ell_{{\scriptscriptstyle L}}\right),\left(\Delta\ell_{{\scriptscriptstyle R}},\ell_{{\scriptscriptstyle L}}\right)\right\} $.
Writing ${\cal A}_{j}=\alpha_{j}\ell$ with $\alpha_{j}$ being some
fixed ratios, we are interested in the leading-order behavior as $\ell\to\infty$.
Again defining the variables $\left\{ \zeta_{j}\right\} $ as in Eq.~(\ref{eq:SPA-Change-of-variables}),
we arrive at the crucial observation that unless ${\cal A}_{1}={\cal A}_{2}=\ldots={\cal A}_{p}$
(and hence also ${\cal B}_{1}={\cal B}_{2}=\ldots={\cal B}_{p}$),
we cannot find in the exponent a $\left(2p-2\right)$-variable function
with a stationary point as before, regardless of the values of $\overrightarrow{\tau},\overrightarrow{\sigma}$.
Leading-order contributions will therefore arise only from terms where
${\cal A}_{1}={\cal A}_{2}=\ldots={\cal A}_{p}$. We can thus conclude
that
\begin{align}
{\rm Tr}\!\left[\left(C_{A}\right)^{p}\right] & \sim\int_{\left[-k_{F,R},k_{F,L}\right]^{p}}\frac{d^{p}k}{\left(2\pi\right)^{p}}\prod_{j=1}^{p}\left\{ \sum_{m=m_{0}+d_{L}+1}^{m_{0}+d_{R}}u_{-m}\!\left(k_{j-1}\right)u_{-m}^{*}\!\left(k_{j}\right)\right\} \nonumber \\
 & +\int_{\left[-k_{F,R},k_{F,L}\right]^{p}}\frac{d^{p}k}{\left(2\pi\right)^{p}}\prod_{j=1}^{p}\left\{ \sum_{m=m_{0}+d_{{\scriptscriptstyle L}}+\ell_{{\scriptscriptstyle L}}+1}^{m_{0}+d_{R}+\ell_{{\scriptscriptstyle R}}}u_{m}\!\left(k_{j-1}\right)u_{m}^{*}\!\left(k_{j}\right)\right\} +{\cal M}^{\left(p\right)},\label{eq:Moments-as-integrals-disjoint-subsystem}
\end{align}
where we defined
\begin{equation}
{\cal M}^{\left(p\right)}=\int_{\left[-k_{F,R},k_{F,L}\right]^{p}}\frac{d^{p}k}{\left(2\pi\right)^{p}}\prod_{j=1}^{p}\left\{ \sum_{m=m_{0}+d_{R}+1}^{m_{0}+d_{L}+\ell_{L}}\left[u_{m}\!\left(k_{j-1}\right)u_{m}^{*}\!\left(k_{j}\right)+u_{-m}\!\left(k_{j-1}\right)u_{-m}^{*}\!\left(k_{j}\right)\right]\right\} .
\end{equation}

The first two integrals in Eq.~(\ref{eq:Moments-as-integrals-disjoint-subsystem})
can be treated in the same way in which the equivalent integrals were
treated in the cases of the connected subsystems. What therefore remains
to be done is to treat ${\cal M}^{\left(p\right)}$. We define $\widetilde{{\cal W}}\!\left(x\right)={\cal W}_{{\scriptscriptstyle L}}\!\left(x\right)e^{2ix\Delta\ell_{{\scriptscriptstyle L}}}$
in order to simplify the notation. In analogy to Eqs.~(\ref{eq:Trace-moment-before-SPA})
and (\ref{eq:Xi-domains-SPA}), we may discard terms that have no
leading-order contribution to ${\cal M}^{\left(p\right)}$ and write
\begin{equation}
{\cal M}^{\left(p\right)}\sim\left(\frac{\ell_{{\rm mirror}}}{4\pi}\right)^{p}\int_{\left[-k_{F,R},k_{F,L}\right]^{p}}d^{p}k\sum_{\overrightarrow{a}\in\left\{ 0,1\right\} ^{\otimes p}}\prod_{j=1}^{p}\left[\widetilde{\Xi}^{a_{j-1}a_{j}}\!\left(k_{a_{j-1}},k_{a_{j}}\right)\Theta\!\left(k_{a_{j}}\right)\right],\label{eq:Initial-SPA-mirrored-sites}
\end{equation}
where
\begin{align}
\widetilde{\Xi}^{00}\!\left(k_{j-1},k_{j}\right) & =\left[t_{{\scriptscriptstyle L}}\!\left(\left|k_{j-1}\right|\right)t_{{\scriptscriptstyle L}}^{*}\!\left(\left|k_{j}\right|\right)+r_{{\scriptscriptstyle L}}\!\left(\left|k_{j-1}\right|\right)r_{{\scriptscriptstyle L}}^{*}\!\left(\left|k_{j}\right|\right)\right]\widetilde{{\cal W}}\!\left(\frac{k_{j-1}-k_{j}}{2}\right)\int_{-1}^{1}d\xi\,e^{\frac{i}{2}\ell_{{\rm mirror}}\left(k_{j-1}-k_{j}\right)\left(\xi+1\right)}\nonumber \\
 & +\widetilde{{\cal W}}\!\left(\frac{k_{j}-k_{j-1}}{2}\right)\int_{-1}^{1}d\xi\,e^{\frac{i}{2}\ell_{{\rm mirror}}\left(k_{j}-k_{j-1}\right)\left(\xi+1\right)},\nonumber \\
\widetilde{\Xi}^{11}\!\left(k_{j-1},k_{j}\right) & =\left[t_{{\scriptscriptstyle R}}\!\left(\left|k_{j-1}\right|\right)t_{{\scriptscriptstyle R}}^{*}\!\left(\left|k_{j}\right|\right)+r_{{\scriptscriptstyle R}}\!\left(\left|k_{j-1}\right|\right)r_{{\scriptscriptstyle R}}^{*}\!\left(\left|k_{j}\right|\right)\right]\widetilde{{\cal W}}\!\left(\frac{k_{j-1}-k_{j}}{2}\right)\int_{-1}^{1}d\xi\,e^{\frac{i}{2}\ell_{{\rm mirror}}\left(k_{j-1}-k_{j}\right)\left(\xi+1\right)}\nonumber \\
 & +\widetilde{{\cal W}}\!\left(\frac{k_{j}-k_{j-1}}{2}\right)\int_{-1}^{1}d\xi\,e^{\frac{i}{2}\ell_{{\rm mirror}}\left(k_{j}-k_{j-1}\right)\left(\xi+1\right)},\nonumber \\
\widetilde{\Xi}^{01}\!\left(k_{j-1},k_{j}\right) & =\left[t_{{\scriptscriptstyle L}}\!\left(\left|k_{j-1}\right|\right)r_{{\scriptscriptstyle R}}^{*}\!\left(\left|k_{j}\right|\right)+r_{{\scriptscriptstyle L}}\!\left(\left|k_{j-1}\right|\right)t_{{\scriptscriptstyle R}}^{*}\!\left(\left|k_{j}\right|\right)\right]\widetilde{{\cal W}}\!\left(\frac{k_{j-1}-k_{j}}{2}\right)\int_{-1}^{1}d\xi\,e^{\frac{i}{2}\ell_{{\rm mirror}}\left(k_{j-1}-k_{j}\right)\left(\xi+1\right)},
\end{align}
and $\widetilde{\Xi}^{10}\!\left(k_{j-1},k_{j}\right)=\widetilde{\Xi}^{01}\!\left(k_{j},k_{j-1}\right)^{*}$.
Applying the SPA through the same procedure as before, while recalling
the unitarity of the scattering matrix in Eq.~(\ref{eq:Scattering_matrix}),
we then obtain
\begin{equation}
{\cal M}^{\left(p\right)}\sim\frac{k_{{\scriptscriptstyle F,L}}+k_{{\scriptscriptstyle F,R}}}{\pi}\ell_{{\rm mirror}},\label{eq:Mirrored-contribution}
\end{equation}
so that in total, 
\begin{equation}
{\rm Tr}\!\left[\left(C_{A}\right)^{p}\right]\sim\begin{cases}
\frac{k_{{\scriptscriptstyle F,L}}+k_{{\scriptscriptstyle F,R}}}{2\pi}\left(\ell_{{\scriptscriptstyle L}}+\ell_{{\scriptscriptstyle R}}\right)+\Delta\ell_{{\scriptscriptstyle L}}\int_{k_{F,R}}^{k_{F,L}}\frac{dk}{2\pi}\left({\cal R}\!\left(k\right)\right)^{p}+\Delta\ell_{{\scriptscriptstyle R}}\int_{k_{F,R}}^{k_{F,L}}\frac{dk}{2\pi}\left[\left({\cal T}\!\left(k\right)\right)^{p}-1\right] & k_{{\scriptscriptstyle F,L}}>k_{{\scriptscriptstyle F,R}},\\
\frac{k_{{\scriptscriptstyle F,L}}+k_{{\scriptscriptstyle F,R}}}{2\pi}\left(\ell_{{\scriptscriptstyle L}}+\ell_{{\scriptscriptstyle R}}\right)+\Delta\ell_{{\scriptscriptstyle L}}\int_{k_{F,L}}^{k_{F,R}}\frac{dk}{2\pi}\left[\left({\cal T}\!\left(k\right)\right)^{p}-1\right]+\Delta\ell_{{\scriptscriptstyle R}}\int_{k_{F,L}}^{k_{F,R}}\frac{dk}{2\pi}\left({\cal R}\!\left(k\right)\right)^{p} & k_{{\scriptscriptstyle F,L}}<k_{{\scriptscriptstyle F,R}}.
\end{cases}\label{eq:Moment-asymptotics-A}
\end{equation}

\subsubsection{Asymptotics of the R\'enyi entropies}

Finally, we use the asymptotics of the moments in Eqs.~(\ref{eq:Moment-asymptotics-A_R}),
(\ref{eq:Moment-asymptotics-A_L}) and (\ref{eq:Moment-asymptotics-A})
to derive the R\'enyi entropies of the subsystems of interest. In
particular, we observe that the terms comprising the series expansion
in Eq.~(\ref{eq:Renyi-from-correlations}) follow the asymptotic
scaling
\begin{align}
{\rm Tr}\!\left[\left\{ \left(C_{A_{i}}\right)^{n}+\left(\mathbb{I}-C_{A_{i}}\right)^{n}-\mathbb{I}\right\} ^{s}\right] & \sim\ell_{i}\int_{k_{-}}^{k_{+}}\frac{dk}{2\pi}\left\{ \left({\cal T}\!\left(k\right)\right)^{n}+\left({\cal R}\!\left(k\right)\right)^{n}-1\right\} ^{s},\nonumber \\
{\rm Tr}\!\left[\left\{ \left(C_{A}\right)^{n}+\left(\mathbb{I}-C_{A}\right)^{n}-\mathbb{I}\right\} ^{s}\right] & \sim\left(\Delta\ell_{{\scriptscriptstyle L}}+\Delta\ell_{{\scriptscriptstyle R}}\right)\int_{k_{-}}^{k_{+}}\frac{dk}{2\pi}\left\{ \left({\cal T}\!\left(k\right)\right)^{n}+\left({\cal R}\!\left(k\right)\right)^{n}-1\right\} ^{s}.
\end{align}
This yields Eq.~(\ref{eq:Renyi-entropies-asymptotics}), which is
true for any $d_{{\scriptscriptstyle L}}$ and $d_{{\scriptscriptstyle R}}$.

\section{Calculation of the fermionic negativity\label{sec:Calculation-of-negativity}}

In this appendix we summarize the derivation of the result for the
fermionic negativity ${\cal E}$ between $A_{{\scriptscriptstyle L}}$
and $A_{{\scriptscriptstyle R}}$. The analytical method we employed
is similar to that used in Appendix \ref{sec:Calculation-of-Renyi-entropies}
for the calculation of the R\'enyi entropies, as we explain below.

The fermionic negativity ${\cal E}$ can be obtained from the R\'enyi
negativities
\begin{equation}
{\cal E}_{n}=\ln{\rm Tr}\!\left[\left(\left(\widetilde{\rho}_{{\scriptscriptstyle A}}\right)^{\dagger}\widetilde{\rho}_{{\scriptscriptstyle A}}\right)^{n/2}\right],
\end{equation}
by evaluating ${\cal E}_{n}$ at even values of $n$ and performing
an analytic continuation to $n=1$. ${\cal E}_{n}$ can be written
in terms of the restricted correlation matrix $C_{A}$ and a transformed
correlation matrix $C_{\Xi}$, facilitating a significant simplification
of the calculation as in the case of the R\'enyi entropies. We write
\begin{equation}
C_{A}=\left(\begin{array}{cc}
C_{A_{L}} & C_{{\scriptscriptstyle LR}}\\
C_{{\scriptscriptstyle RL}} & C_{A_{R}}
\end{array}\right),
\end{equation}
where the matrices $C_{{\scriptscriptstyle LR}}$ and $C_{{\scriptscriptstyle RL}}=\left(C_{{\scriptscriptstyle LR}}\right)^{\dagger}$
represent two-point correlations between a site in $A_{{\scriptscriptstyle L}}$
and another in $A_{{\scriptscriptstyle R}}$, and we define 
\begin{equation}
C_{\Xi}=\frac{1}{2}\left[\mathbb{I}-\left(\mathbb{I}+\Gamma_{+}\Gamma_{-}\right)^{-1}\left(\Gamma_{+}+\Gamma_{-}\right)\right],
\end{equation}
where
\begin{equation}
\Gamma_{\pm}=\left(\begin{array}{cc}
2C_{A_{L}}-\mathbb{I}_{\ell_{L}} & \mp2iC_{{\scriptscriptstyle LR}}\\
\mp2iC_{{\scriptscriptstyle RL}} & \mathbb{I}_{\ell_{R}}-2C_{A_{R}}
\end{array}\right).
\end{equation}
The R\'enyi negativities can then be written as~\citep{PhysRevB.95.165101,PhysRevB.97.165123,Shapourian_2019}
\begin{equation}
{\cal E}_{n}=\ln\det\!\left[\left(C_{\Xi}\right)^{n/2}+\left(\mathbb{I}-C_{\Xi}\right)^{n/2}\right]+\frac{n}{2}\ln\det\!\left[\left(C_{A}\right)^{2}+\left(\mathbb{I}-C_{A}\right)^{2}\right].\label{eq:Negativity-from-correlations}
\end{equation}

We now define the polynomials 
\begin{equation}
p_{n}\!\left(z\right)=z^{n}+\left(1-z\right)^{n}=\prod_{\gamma=-\frac{n-1}{2}}^{\frac{n-1}{2}}\left(1-\frac{z}{z_{\gamma}}\right)\label{eq:MI-polynomial-definition}
\end{equation}
and
\begin{equation}
\tilde{p}_{n}\!\left(z\right)=z^{n/2}+\left(1-z\right)^{n/2}=\prod_{\gamma=1/2}^{\frac{n-1}{2}}\left(1-\frac{z}{\tilde{z}_{\gamma}}\right)\label{eq:Negativity-polynomial-definition}
\end{equation}
for any even integer $n$. Here $\left\{ z_{\gamma}\right\} $ and
$\left\{ \tilde{z}_{\gamma}\right\} $ are, respectively, the roots
of $p_{n}$ and $\tilde{p}_{n}$, and they satisfy
\begin{align}
\left(z_{\gamma}\right)^{-1} & =1-e^{2\pi i\gamma/n},\,\,\,\,\,\gamma=-\frac{n-1}{2},-\frac{n-3}{2},\ldots,\frac{n-1}{2},\nonumber \\
\left(\tilde{z}_{\gamma}\right)^{-1} & =\frac{e^{2\pi i\gamma/n}+e^{-2\pi i\gamma/n}}{e^{2\pi i\gamma/n}},\,\,\,\,\,\gamma=\frac{1}{2},\frac{3}{2},\ldots,\frac{n-1}{2}.\label{eq:Polynomial-roots}
\end{align}
Note that $p_{n}$ has $n$ different roots, while $\tilde{p}_{n}$
has $n/2$ roots if $n=0\!\!\!\mod\!4$, and $n/2-1$ roots if $n=2\!\!\!\mod\!4$;
in the latter case the missing root corresponds to the index $\gamma=n/4$,
for which $1/\tilde{z}_{\gamma}=0$. Additionally, we recognize that
$\det\!\left[\mathbb{I}+\Gamma_{+}\Gamma_{-}\right]=\det\!\left[\mathbb{I}+\left(\mathbb{I}-2C_{A}\right)^{2}\right]$,
so that using the definition of $\tilde{p}_{n}$ we may write the
R\'enyi negativities in Eq.~(\ref{eq:Negativity-from-correlations})
as
\begin{equation}
{\cal E}_{n}=\ln\det\!\left[\prod_{\gamma=1/2}^{\frac{n-1}{2}}\left[\frac{\mathbb{I}+\Gamma_{+}\Gamma_{-}}{2}-\frac{\left(\mathbb{I}-\Gamma_{+}\right)\left(\mathbb{I}-\Gamma_{-}\right)}{4\tilde{z}_{\gamma}}\right]\right].\label{eq:Renyi-negativity-gamma-decomposition}
\end{equation}

Now, if we define the modified correlation matrices 
\begin{align}
C_{\gamma} & =\left(\begin{array}{cc}
\left(1-e^{\frac{2\pi i\gamma}{n}}\right)\mathbb{I}_{\ell_{L}} & 0\\
0 & \left(1+e^{\frac{-2\pi i\gamma}{n}}\right)\mathbb{I}_{\ell_{R}}
\end{array}\right)C_{A}=\left(\begin{array}{cc}
\left(1-e^{\frac{2\pi i\gamma}{n}}\right)C_{A_{L}} & \left(1-e^{\frac{2\pi i\gamma}{n}}\right)C_{{\scriptscriptstyle LR}}\\
\left(1+e^{\frac{-2\pi i\gamma}{n}}\right)C_{{\scriptscriptstyle RL}} & \left(1+e^{\frac{-2\pi i\gamma}{n}}\right)C_{A_{R}}
\end{array}\right),\nonumber \\
C_{\gamma}' & =C_{A}\left(\begin{array}{cc}
\left(1-e^{\frac{2\pi i\gamma}{n}}\right)\mathbb{I}_{\ell_{L}} & 0\\
0 & \left(1+e^{\frac{-2\pi i\gamma}{n}}\right)\mathbb{I}_{\ell_{R}}
\end{array}\right)=\left(\begin{array}{cc}
\left(1-e^{\frac{2\pi i\gamma}{n}}\right)C_{A_{L}} & \left(1+e^{\frac{-2\pi i\gamma}{n}}\right)C_{{\scriptscriptstyle LR}}\\
\left(1-e^{\frac{2\pi i\gamma}{n}}\right)C_{{\scriptscriptstyle RL}} & \left(1+e^{\frac{-2\pi i\gamma}{n}}\right)C_{A_{R}}
\end{array}\right),
\end{align}
one may check that
\begin{align}
\frac{\mathbb{I}+\Gamma_{+}\Gamma_{-}}{2}-\frac{\left(\mathbb{I}-\Gamma_{+}\right)\left(\mathbb{I}-\Gamma_{-}\right)}{4\tilde{z}_{\gamma}} & =\left(\begin{array}{cc}
i\mathbb{I}_{\ell_{L}} & 0\\
0 & \mathbb{I}_{\ell_{R}}
\end{array}\right)\left(\mathbb{I}-C_{\gamma}'\right)\left(\begin{array}{cc}
-e^{\frac{-4\pi i\gamma}{n}}\mathbb{I}_{\ell_{L}} & 0\\
0 & \mathbb{I}_{\ell_{R}}
\end{array}\right)\left(\mathbb{I}-C_{\gamma-\frac{n}{2}}\right)\left(\begin{array}{cc}
-i\mathbb{I}_{\ell_{L}} & 0\\
0 & \mathbb{I}_{\ell_{R}}
\end{array}\right).\label{eq:Renyi-negativity-gamma-components}
\end{align}
By substituting Eq.~(\ref{eq:Renyi-negativity-gamma-components})
into Eq.~(\ref{eq:Renyi-negativity-gamma-decomposition}), and recognizing
that $\prod_{\gamma=1/2}^{\left(n-1\right)/2}\left(-e^{\frac{-4\pi i\gamma}{n}}\right)=1$,
we arrive at the result
\begin{align}
{\cal E}_{n} & =\ln\det\!\left[\prod_{\gamma=1/2}^{\frac{n-1}{2}}\left(\mathbb{I}-C_{\gamma}'\right)\left(\mathbb{I}-C_{-\gamma}\right)\right]={\rm Tr}\ln\!\left[\prod_{\gamma=-\frac{n-1}{2}}^{\frac{n-1}{2}}\left(\mathbb{I}-C_{\gamma}\right)\right],
\end{align}
which then yields the series expansion of ${\cal E}_{n}$ reported
in Eq.~(\ref{eq:Renyi_negativity_series_expansion}). As explained
in Subsec.~\ref{subsec:Derivation_of_negativity}, writing this series
expansion reduces the calculation to that of terms of the form ${\rm Tr}\!\left[C_{\gamma_{1}}C_{\gamma_{2}}\ldots C_{\gamma_{p}}\right]$,
corresponding to the general integral expression in Eq.~(\ref{eq:Renyi-negativity-integral-decomposition}).

We proceed by applying the SPA to the integrals appearing in Eq.~(\ref{eq:Renyi-negativity-integral-decomposition}).
For concreteness we again assume that $d_{{\scriptscriptstyle L}}<d_{{\scriptscriptstyle R}}<d_{{\scriptscriptstyle L}}+\ell_{{\scriptscriptstyle L}}<d_{{\scriptscriptstyle R}}+\ell_{{\scriptscriptstyle R}}$.
The same argument that led to Eq.~(\ref{eq:Moments-as-integrals-disjoint-subsystem})
allows us to separate the integral into independent leading-order
contributions arising from mirrored and unmirrored sites. Namely,
to the linear order in $\Delta\ell_{{\scriptscriptstyle L}}$, $\Delta\ell_{{\scriptscriptstyle R}}$
and $\ell_{{\rm mirror}}$, we have
\begin{align}
{\rm Tr}\!\left[C_{\gamma_{1}}\ldots C_{\gamma_{p}}\right] & \sim\int_{\left[-k_{F,R},k_{F,L}\right]^{p}}\!\!\frac{d^{p}k}{\left(2\pi\right)^{p}}\prod_{j=1}^{p}\!\left[\left(1-e^{\frac{2\pi i\gamma_{j}}{n}}\right)\sum_{m=m_{0}+d_{L}+1}^{m_{0}+d_{R}}\!u_{-m}\!\left(k_{j-1}\right)u_{-m}^{*}\!\left(k_{j}\right)\right]\nonumber \\
 & +\int_{\left[-k_{F,R},k_{F,L}\right]^{p}}\!\!\frac{d^{p}k}{\left(2\pi\right)^{p}}\prod_{j=1}^{p}\!\left[\left(1+e^{\frac{-2\pi i\gamma_{j}}{n}}\right)\sum_{m=m_{0}+d_{{\scriptscriptstyle L}}+\ell_{{\scriptscriptstyle L}}+1}^{m_{0}+d_{R}+\ell_{{\scriptscriptstyle R}}}\!u_{m}\!\left(k_{j-1}\right)u_{m}^{*}\!\left(k_{j}\right)\right]+{\cal M}_{\gamma_{1}\ldots\gamma_{p}},\label{eq:Renyi-negativities-as-integrals}
\end{align}
where we defined
\begin{equation}
{\cal M}_{\gamma_{1}\ldots\gamma_{p}}=\int_{\left[-k_{F,R},k_{F,L}\right]^{p}}\!\!\frac{d^{p}k}{\left(2\pi\right)^{p}}\prod_{j=1}^{p}\left\{ \sum_{m=m_{0}+d_{R}+1}^{m_{0}+d_{L}+\ell_{L}}\left[\left(1-e^{\frac{2\pi i\gamma_{j}}{n}}\right)u_{-m}\!\left(k_{j-1}\right)u_{-m}^{*}\!\left(k_{j}\right)+\left(1+e^{\frac{-2\pi i\gamma_{j}}{n}}\right)u_{m}\!\left(k_{j-1}\right)u_{m}^{*}\!\left(k_{j}\right)\right]\right\} .
\end{equation}
The first two integrals in Eq.~(\ref{eq:Renyi-negativities-as-integrals})
have already been treated using the SPA in Appendix~\ref{sec:Calculation-of-Renyi-entropies},
as they arise (up to a multiplicative constant) in the calculation
of the R\'enyi entropies (see Eqs.~(\ref{eq:Moment-asymptotics-A_R})
and (\ref{eq:Moment-asymptotics-A_L})). Assuming for concreteness
that $k_{{\scriptscriptstyle F,L}}>k_{{\scriptscriptstyle F,R}}$,
we may therefore write
\begin{align}
{\rm Tr}\!\left[C_{\gamma_{1}}\ldots C_{\gamma_{p}}\right] & \sim\Delta\ell_{{\scriptscriptstyle L}}\left[\frac{k_{{\scriptscriptstyle F,R}}+k_{{\scriptscriptstyle F,L}}}{2\pi}+\int_{k_{F,R}}^{k_{F,L}}\frac{dk}{2\pi}\left({\cal R}\!\left(k\right)\right)^{p}\right]\prod_{j=1}^{p}\left(1-e^{\frac{2\pi i\gamma_{j}}{n}}\right)\nonumber \\
 & +\Delta\ell_{{\scriptscriptstyle R}}\left[\frac{k_{{\scriptscriptstyle F,R}}}{\pi}+\int_{k_{F,R}}^{k_{F,L}}\frac{dk}{2\pi}\left({\cal T}\!\left(k\right)\right)^{p}\right]\prod_{j=1}^{p}\left(1+e^{\frac{-2\pi i\gamma_{j}}{n}}\right)+{\cal M}_{\gamma_{1}\ldots\gamma_{p}}.\label{eq:Negativity-product-asymptotics-unmirrored}
\end{align}

Let us now address the asymptotics of ${\cal M}_{\gamma_{1}\ldots\gamma_{p}}$,
to the linear order in $\ell_{{\rm mirror}}$. Repeating the argument
in Appendix \ref{sec:Calculation-of-Renyi-entropies} leading to Eq.~(\ref{eq:Xi-domains-SPA}),
which was also used to obtain Eq.~(\ref{eq:Initial-SPA-mirrored-sites}),
we use the SPA to discard terms with no leading-order contribution
and write
\begin{equation}
{\cal M}_{\gamma_{1}\ldots\gamma_{p}}\sim\left(\frac{\ell_{{\rm mirror}}}{4\pi}\right)^{p}\int_{\left[-k_{F,R},k_{F,L}\right]^{p}}d^{p}k\sum_{\overrightarrow{a}\in\left\{ 0,1\right\} ^{\otimes p}}\prod_{j=1}^{p}\left[\widetilde{\Xi}_{\gamma_{j}}^{a_{j-1}a_{j}}\!\left(k_{a_{j-1}},k_{a_{j}}\right)\Theta\!\left(k_{a_{j}}\right)\right],\label{eq:Initial-SPA-mirrored-sites-Negativity}
\end{equation}
with
\begin{align}
\widetilde{\Xi}_{\gamma}^{00}\!\left(k_{j-1},k_{j}\right) & =\left[\left(1+e^{\frac{-2\pi i\gamma}{n}}\right)t_{{\scriptscriptstyle L}}\!\left(\left|k_{j-1}\right|\right)t_{{\scriptscriptstyle L}}^{*}\!\left(\left|k_{j}\right|\right)+\left(1-e^{\frac{2\pi i\gamma}{n}}\right)r_{{\scriptscriptstyle L}}\!\left(\left|k_{j-1}\right|\right)r_{{\scriptscriptstyle L}}^{*}\!\left(\left|k_{j}\right|\right)\right]\nonumber \\
 & \times\widetilde{{\cal W}}\!\left(\frac{k_{j-1}-k_{j}}{2}\right)\int_{-1}^{1}d\xi\,e^{\frac{i}{2}\ell_{{\rm mirror}}\left(k_{j-1}-k_{j}\right)\left(\xi+1\right)}+\left(1-e^{\frac{2\pi i\gamma}{n}}\right)\widetilde{{\cal W}}\!\left(\frac{k_{j}-k_{j-1}}{2}\right)\int_{-1}^{1}d\xi\,e^{\frac{i}{2}\ell_{{\rm mirror}}\left(k_{j}-k_{j-1}\right)\left(\xi+1\right)},\nonumber \\
\widetilde{\Xi}_{\gamma}^{11}\!\left(k_{j-1},k_{j}\right) & =\left[\left(1-e^{\frac{2\pi i\gamma}{n}}\right)t_{{\scriptscriptstyle R}}\!\left(\left|k_{j-1}\right|\right)t_{{\scriptscriptstyle R}}^{*}\!\left(\left|k_{j}\right|\right)+\left(1+e^{\frac{-2\pi i\gamma}{n}}\right)r_{{\scriptscriptstyle R}}\!\left(\left|k_{j-1}\right|\right)r_{{\scriptscriptstyle R}}^{*}\!\left(\left|k_{j}\right|\right)\right]\nonumber \\
 & \times\widetilde{{\cal W}}\!\left(\frac{k_{j-1}-k_{j}}{2}\right)\int_{-1}^{1}d\xi\,e^{\frac{i}{2}\ell_{{\rm mirror}}\left(k_{j-1}-k_{j}\right)\left(\xi+1\right)}+\left(1+e^{\frac{-2\pi i\gamma}{n}}\right)\widetilde{{\cal W}}\!\left(\frac{k_{j}-k_{j-1}}{2}\right)\int_{-1}^{1}d\xi\,e^{\frac{i}{2}\ell_{{\rm mirror}}\left(k_{j}-k_{j-1}\right)\left(\xi+1\right)},\nonumber \\
\widetilde{\Xi}_{\gamma}^{01}\!\left(k_{j-1},k_{j}\right) & =\left[\left(1+e^{\frac{-2\pi i\gamma}{n}}\right)t_{{\scriptscriptstyle L}}\!\left(\left|k_{j-1}\right|\right)r_{{\scriptscriptstyle R}}^{*}\!\left(\left|k_{j}\right|\right)+\left(1-e^{\frac{2\pi i\gamma}{n}}\right)r_{{\scriptscriptstyle L}}\!\left(\left|k_{j-1}\right|\right)t_{{\scriptscriptstyle R}}^{*}\!\left(\left|k_{j}\right|\right)\right]\nonumber \\
 & \times\widetilde{{\cal W}}\!\left(\frac{k_{j-1}-k_{j}}{2}\right)\int_{-1}^{1}d\xi\,e^{\frac{i}{2}\ell_{{\rm mirror}}\left(k_{j-1}-k_{j}\right)\left(\xi+1\right)},\nonumber \\
\widetilde{\Xi}_{\gamma}^{10}\!\left(k_{j-1},k_{j}\right) & =\left[\left(1+e^{\frac{-2\pi i\gamma}{n}}\right)r_{{\scriptscriptstyle R}}\!\left(\left|k_{j-1}\right|\right)t_{{\scriptscriptstyle L}}^{*}\!\left(\left|k_{j}\right|\right)+\left(1-e^{\frac{2\pi i\gamma}{n}}\right)t_{{\scriptscriptstyle R}}\!\left(\left|k_{j-1}\right|\right)r_{{\scriptscriptstyle L}}^{*}\!\left(\left|k_{j}\right|\right)\right]\nonumber \\
 & \times\widetilde{{\cal W}}\!\left(\frac{k_{j-1}-k_{j}}{2}\right)\int_{-1}^{1}d\xi\,e^{\frac{i}{2}\ell_{{\rm mirror}}\left(k_{j-1}-k_{j}\right)\left(\xi+1\right)}.
\end{align}
 According to Eq.~(\ref{eq:Leading-order-contributing-integrals}),
the SPA imposes the restriction $\left|k_{1}\right|=\left|k_{2}\right|=\ldots=\left|k_{p}\right|$
on integrals contributing to the leading order; since in Eq.~(\ref{eq:Initial-SPA-mirrored-sites-Negativity})
the integration subdomain with $k_{{\scriptscriptstyle F,R}}<\left|k_{j}\right|<k_{{\scriptscriptstyle F,L}}$
is limited to $k_{j}>0$, ${\cal M}_{\gamma_{1}\ldots\gamma_{p}}$
is split into two independent contributions,
\begin{align}
{\cal M}_{\gamma_{1}\ldots\gamma_{p}} & \sim\left(\frac{\ell_{{\rm mirror}}}{4\pi}\right)^{p}\int_{\left[-k_{F,R},k_{F,R}\right]^{p}}d^{p}k\sum_{\overrightarrow{a}\in\left\{ 0,1\right\} ^{\otimes p}}\prod_{j=1}^{p}\left[\widetilde{\Xi}_{\gamma_{j}}^{a_{j-1}a_{j}}\!\left(k_{a_{j-1}},k_{a_{j}}\right)\Theta\!\left(k_{a_{j}}\right)\right]\nonumber \\
 & +\left(\frac{\ell_{{\rm mirror}}}{4\pi}\right)^{p}\int_{\left[k_{F,R},k_{F,L}\right]^{p}}d^{p}k\prod_{j=1}^{p}\widetilde{\Xi}_{\gamma_{j}}^{00}\!\left(k_{j-1},k_{j}\right).\label{eq:SPA-products-mirrored-sites}
\end{align}
The asymptotics of both integrals can be estimated using the SPA procedure
explained before. When applied to the first integral (which corresponds
to an equilibrium scenario), this procedure yields
\begin{equation}
\left(\frac{\ell_{{\rm mirror}}}{4\pi}\right)^{p}\!\!\!\int_{\left[-k_{F,R},k_{F,R}\right]^{p}}\!\!\!\!\!\!\!d^{p}k\sum_{\overrightarrow{a}\in\left\{ 0,1\right\} ^{\otimes p}}\prod_{j=1}^{p}\!\left[\widetilde{\Xi}_{\gamma_{j}}^{a_{j-1}a_{j}}\!\left(k_{a_{j-1}},k_{a_{j}}\right)\Theta\!\left(k_{a_{j}}\right)\right]\sim\ell_{{\rm mirror}}\frac{k_{{\scriptscriptstyle F,R}}}{\pi}\left\{ \prod_{j=1}^{p}\!\left(1-e^{\frac{2\pi i\gamma_{j}}{n}}\right)+\prod_{j=1}^{p}\!\left(1+e^{\frac{-2\pi i\gamma_{j}}{n}}\right)\right\} ,
\end{equation}
while for the second integral we find
\begin{equation}
\left(\frac{\ell_{{\rm mirror}}}{4\pi}\right)^{p}\!\!\!\int_{\left[k_{F,R},k_{F,L}\right]^{p}}\!\!\!\!\!\!\!d^{p}k\prod_{j=1}^{p}\widetilde{\Xi}_{\gamma_{j}}^{00}\!\left(k_{j-1},k_{j}\right)\sim\ell_{{\rm mirror}}\int_{k_{F,R}}^{k_{F,L}}\frac{dk}{2\pi}\left\{ \prod_{j=1}^{p}\left(1-e^{\frac{2\pi i\gamma_{j}}{n}}\right)+\prod_{j=1}^{p}\left[1-e^{\frac{2\pi i\gamma_{j}}{n}}{\cal R}\!\left(k\right)+e^{\frac{-2\pi i\gamma_{j}}{n}}{\cal T}\!\left(k\right)\right]\right\} .
\end{equation}

Together with Eq.~(\ref{eq:Negativity-product-asymptotics-unmirrored}),
we then have
\begin{align}
{\rm Tr}\left[\left\{ \prod_{\gamma=-\frac{n-1}{2}}^{\frac{n-1}{2}}\left(\mathbb{I}-C_{\gamma}\right)-\mathbb{I}\right\} ^{\!s}\,\right] & \sim\ell_{{\rm mirror}}\int_{k_{F,R}}^{k_{F,L}}\frac{dk}{2\pi}\left\{ \prod_{\gamma=-\frac{n-1}{2}}^{\frac{n-1}{2}}\left(e^{\frac{2\pi i\gamma}{n}}{\cal R}\!\left(k\right)-e^{\frac{-2\pi i\gamma}{n}}{\cal T}\!\left(k\right)\right)-1\right\} ^{\!s}\nonumber \\
 & +\Delta\ell_{{\scriptscriptstyle L}}\int_{k_{F,R}}^{k_{F,L}}\frac{dk}{2\pi}\left\{ \prod_{\gamma=-\frac{n-1}{2}}^{\frac{n-1}{2}}\left(1-\left(1-e^{\frac{2\pi i\gamma}{n}}\right){\cal R}\!\left(k\right)\right)-1\right\} ^{\!s}\nonumber \\
 & +\Delta\ell_{{\scriptscriptstyle R}}\int_{k_{F,R}}^{k_{F,L}}\frac{dk}{2\pi}\left\{ \prod_{\gamma=-\frac{n-1}{2}}^{\frac{n-1}{2}}\left(1-\left(1+e^{\frac{-2\pi i\gamma}{n}}\right){\cal T}\!\left(k\right)\right)-1\right\} ^{\!s}.
\end{align}
Using the decomposition of the polynomials $p_{n}$ and $\tilde{p}_{n}$
in Eqs.~(\ref{eq:MI-polynomial-definition}) and (\ref{eq:Negativity-polynomial-definition}),
we arrive at Eq\@.~(\ref{eq:Renyi-negativity-expansion-terms-asymptotics}),
a result that also captures the case $k_{{\scriptscriptstyle F,L}}<k_{{\scriptscriptstyle F,R}}$
(for which an equivalent derivation applies). By summing the series
in Eq.~(\ref{eq:Renyi_negativity_series_expansion}) and taking the
limit $n\to1$, we then obtain the final result for the leading-order
asymptotics of the fermionic negativity, given by Eq.~(\ref{eq:Negativity_volume_law}).

\begin{figure}
\begin{centering}
\includegraphics[viewport=20bp 0bp 930bp 555bp,clip,width=1\textwidth]{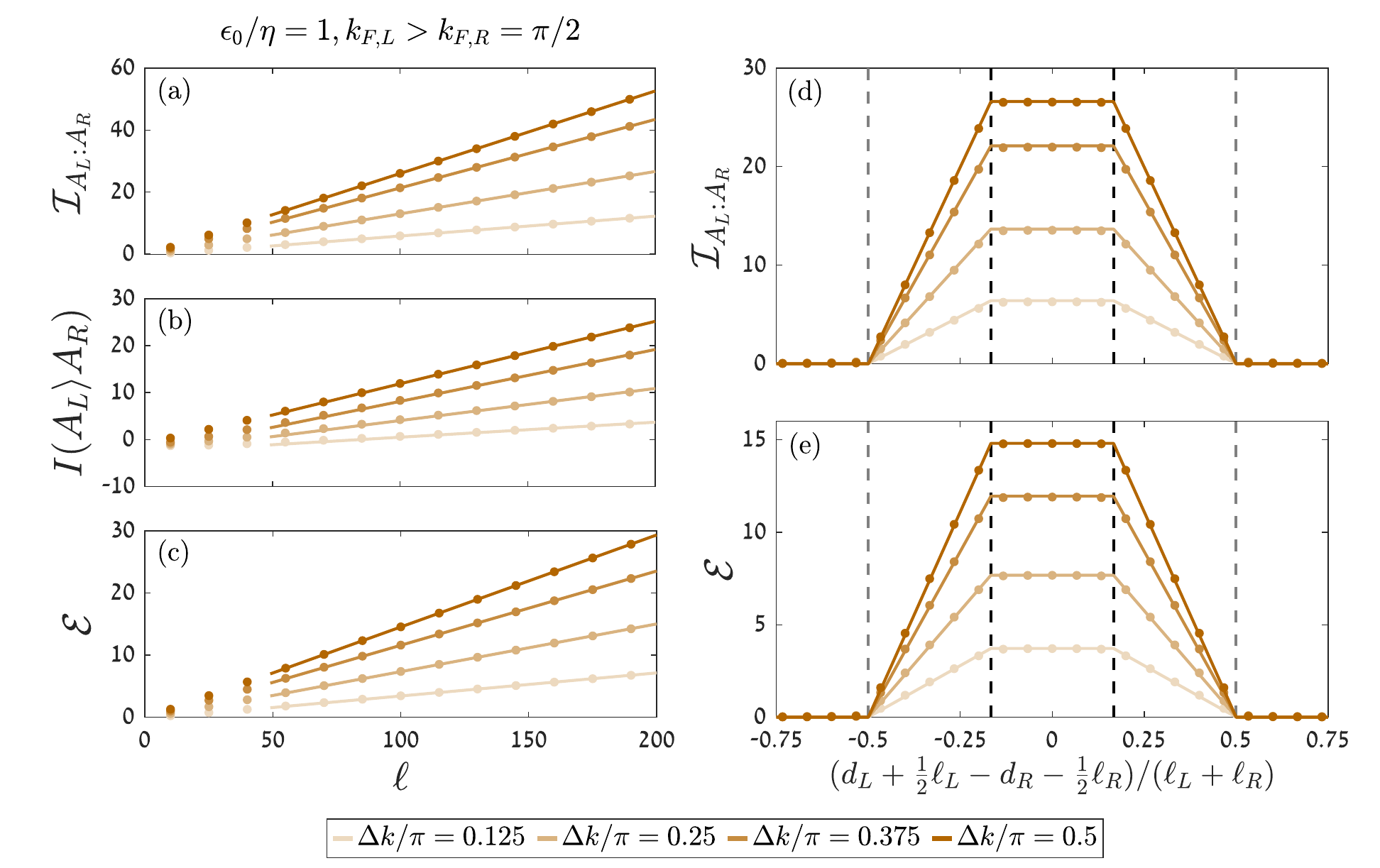}
\par\end{centering}
\caption{\label{fig:Num-comparison-symmetric-bias}The single impurity model:
Scaling of (a) the mutual information, (b) the coherent information,
and (c) the fermionic negativity between subsystems $A_{{\scriptscriptstyle L}}$
and $A_{{\scriptscriptstyle R}}$ for the symmetric case $\ell_{{\scriptscriptstyle L}}=\ell_{{\scriptscriptstyle R}}=\ell$
and $d_{{\scriptscriptstyle L}}=d_{{\scriptscriptstyle R}}$. (d)
The mutual information and (e) the fermionic negativity as a function
of $d_{{\scriptscriptstyle L}}-d_{{\scriptscriptstyle R}}$, when
fixing $\ell_{{\scriptscriptstyle L}}=100$ and $\ell_{{\scriptscriptstyle R}}=200$;
letting $\bar{A}_{{\scriptscriptstyle L}}=\left\{ m|-m\in A_{{\scriptscriptstyle L}}\right\} $
denote the mirror image of $A_{{\scriptscriptstyle L}}$, black dashed
vertical lines mark the boundaries of the domain where $\bar{A}_{{\scriptscriptstyle L}}\subset A_{{\scriptscriptstyle R}}$,
while gray dashed vertical lines mark the boundaries of the domain
where $\bar{A}_{{\scriptscriptstyle L}}\cap A_{{\scriptscriptstyle R}}\protect\neq\phi$.
In all the panels, results are computed in the limit $d_{i}\gg\ell_{i}$;
analytical results (lines) are compared to numerical results (dots)
for different values of the bias $\Delta k=k_{{\scriptscriptstyle F,L}}-k_{{\scriptscriptstyle F,R}}$,
with the lower Fermi momentum fixed at $k_{{\scriptscriptstyle F,R}}=\pi/2$,
and the impurity energy fixed at $\epsilon_{0}=\eta$.}
\end{figure}

\section{Additional numerical tests\label{sec:Additional-numerical-tests}}

Fig.~\ref{fig:Num-Comparison-Symmetric} shows a comparison between
our analytical results and numerical calculations of the different
correlation measures -- MI, CI and fermionic negativity -- assuming
a symmetric configuration of the subsystems ($\ell_{{\scriptscriptstyle L}}=\ell_{{\scriptscriptstyle R}}$
and $d_{{\scriptscriptstyle L}}=d_{{\scriptscriptstyle R}}$); Fig.~\ref{fig:MI-Varying-Distance}
compares analytical and numerical results for the dependence of the
MI and negativity on the positions relative to the scatterer of subsystems
with fixed lengths. All of these results were computed for the single
impurity model described in Sec.~\ref{sec:Asymptotics-of-correlation-measures},
for fixed values of the Fermi momenta and various values of the impurity
energy $\epsilon_{0}$. In Fig.~\ref{fig:Num-comparison-symmetric-bias}
we show similar comparisons, where now the impurity energy is fixed
and results are plotted for various values of the bias, which is another
parameter that influences the asymptotic scaling coefficients (for
a bias that is small enough such that the scattering probabilities
vary negligibly in $\left[k_{-},k_{+}\right]$, the leading volume-law
terms of the correlations measures are linear in the bias). Again,
the good agreement of the analytical calculation with numerics is
clearly evident.

The numerical calculations presented in Figs.~\ref{fig:Num-Comparison-Symmetric}--\ref{fig:Num-comparison-symmetric-bias}
all rely on the direct diagonalization of two-point correlation matrices,
through Eq.~(\ref{eq:Renyi-from-correlations}) (for the MI and CI)
and Eq.~(\ref{eq:Negativity-from-correlations}) (for the negativity).
The entries of these correlation matrices were computed in the limit
$d_{i}/\ell_{i}\to\infty$ (with $d_{{\scriptscriptstyle L}}-d_{{\scriptscriptstyle R}}$
kept fixed), by discarding terms that vanish in this limit according
to the Riemann-Lebesgue lemma, as explained in Appendix \ref{sec:Two-point-correlations}.

In Fig.~\ref{fig:Numerical-distance-dependence} we demonstrate that
the omission of these terms from the correlation matrices indeed captures
the $d_{i}/\ell_{i}\to\infty$ limit of the correlation measures themselves.
For the symmetric case $\ell_{{\scriptscriptstyle L}}=\ell_{{\scriptscriptstyle R}}=\ell$
and $d_{{\scriptscriptstyle L}}=d_{{\scriptscriptstyle R}}=d$, we
let ${\cal I}_{A_{L}:A_{R}}^{\left(d\right)}$ and ${\cal E}^{\left(d\right)}$
denote the MI and negativity, respectively, that were numerically
calculated using correlation matrices with entries given by Eq.~(\ref{eq:Correlation_function_integral}),
while ${\cal I}_{A_{L}:A_{R}}^{\left(\infty\right)}$ and ${\cal E}^{\left(\infty\right)}$
stand, respectively, for the MI and negativity that were numerically
calculated using correlation matrices with entries given by Eq.~(\ref{eq:Correlation-function-large-distance}).

Indeed, the results in Fig.~\ref{fig:Numerical-distance-dependence},
numerically computed for the single impurity model, indicate that
${\cal I}_{A_{L}:A_{R}}^{\left(d\right)}\to{\cal I}_{A_{L}:A_{R}}^{\left(\infty\right)}$
and ${\cal E}^{\left(d\right)}\to{\cal E}^{\left(\infty\right)}$
as $d/\ell\to\infty$. As they converge toward this limit, the correlation
measures exhibit Friedel oscillations, a behavior that was previously
observed for the entanglement entropy of a single subsystem of contiguous
sites~\citep{10.21468/SciPostPhys.11.4.085}. As in that case, the
difference between the average over these oscillations and the $d\to\infty$
limit decays according to a $\propto1/d^{2}$ power law, while the
amplitude of the oscillations decays according to a $\propto1/d$
power law.

\begin{figure}
\begin{centering}
\includegraphics[viewport=90bp 0bp 1280bp 480bp,clip,width=1\textwidth]{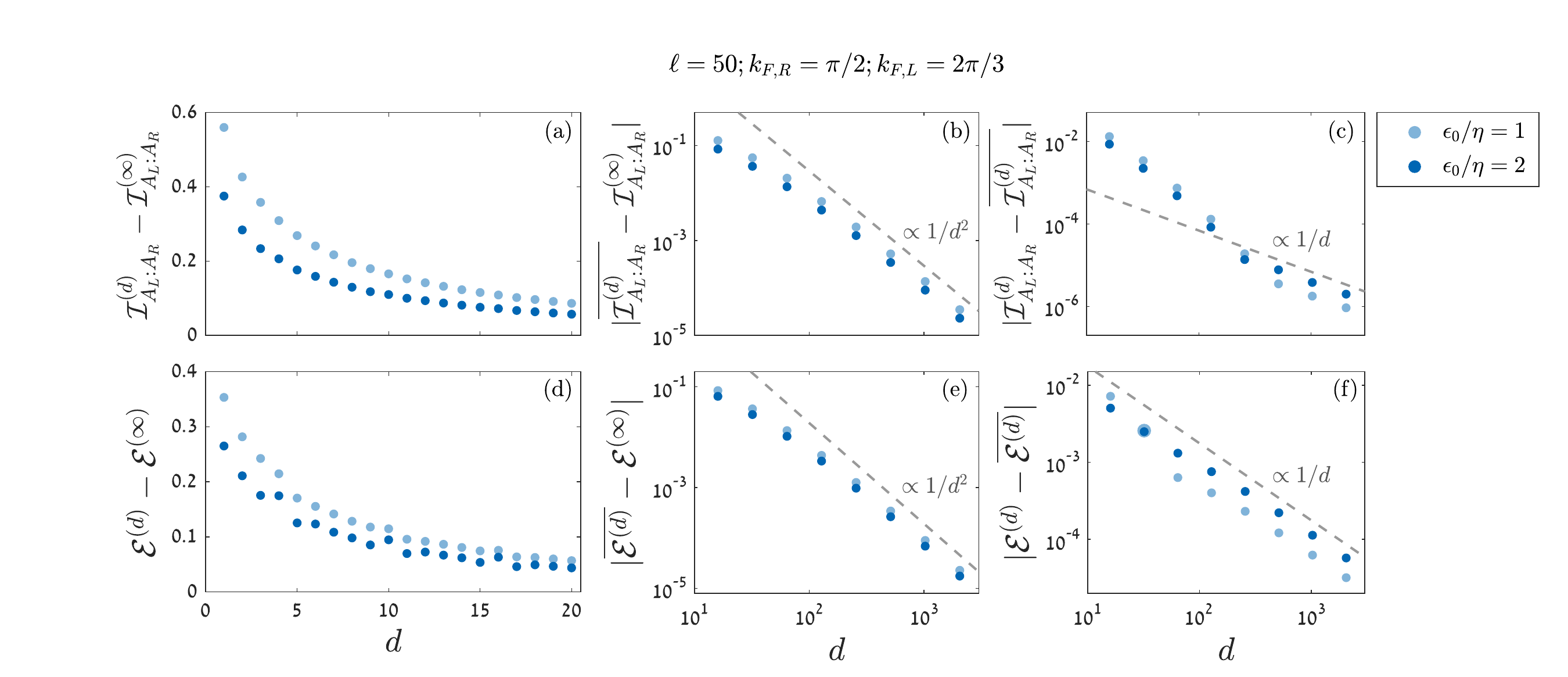}
\par\end{centering}
\caption{\label{fig:Numerical-distance-dependence}The single impurity model:
Comparison between numerical calculations of correlation measures
for the symmetric case with $\ell_{{\scriptscriptstyle L}}=\ell_{{\scriptscriptstyle R}}=\ell=50$
and $d_{{\scriptscriptstyle L}}=d_{{\scriptscriptstyle R}}=d$, for
two values of the impurity energy $\epsilon_{0}$ and with the Fermi
momenta fixed at $k_{{\scriptscriptstyle F,R}}=\pi/2$ and $k_{{\scriptscriptstyle F,L}}=2\pi/3$.
(a) The difference between ${\cal I}_{A_{L}:A_{R}}^{\left(d\right)}$,
the MI computed using the full expressions for the correlation matrices
(Eq.~(\ref{eq:Correlation_function_integral})), and ${\cal I}_{A_{L}:A_{R}}^{\left(\infty\right)}$,
the MI computed from correlation matrices where entries were taken
to the limit $d\to\infty$ (Eq.~(\ref{eq:Correlation-function-large-distance})),
as a function of $d$. (b) The deviation of $\overline{{\cal I}_{A_{L}:A_{R}}^{\left(d\right)}}$,
the average of ${\cal I}_{A_{L}:A_{R}}^{\left(d\right)}$ over Friedel
oscillations, from ${\cal I}_{A_{L}:A_{R}}^{\left(\infty\right)}$;
the dashed gray line emphasizes that, for $d\gg\ell$, the deviation
approaches a $\propto1/d^{2}$ power-law behavior. (c) The amplitude
$\left|{\cal I}_{A_{L}:A_{R}}^{\left(d\right)}-\overline{{\cal I}_{A_{L}:A_{R}}^{\left(d\right)}}\right|$
of the oscillations in the MI; the dashed gray line emphasizes that,
for $d\gg\ell$, the amplitude approaches a $\propto1/d$ power-law
behavior. The bottom panels (d)--(f) present a similar analysis for
the fermionic negativity ${\cal E}$.}
\end{figure}

\bibliography{Extensive_Entanglement_NESS}

\end{document}